\documentclass[aps,prd,twocolumn,showpacs,%
nofootinbib,amsmath,amssymb,amsfonts]{revtex4}
\usepackage{bm} 
\usepackage{bbm} 
\begin{document}
\title{Vector theories in cosmology}

\author{Gilles Esposito-Far\`ese}
\email{gef@iap.fr}
\affiliation{${\mathcal{G}}{\mathbb{R}}
\varepsilon{\mathbb{C}}{\mathcal{O}}$, Institut d'Astrophysique
de Paris, UMR 7095-CNRS, Universit\'e Pierre et Marie
Curie-Paris 6, 98bis boulevard Arago, F-75014 Paris, France}

\author{Cyril Pitrou}
\email{cyril.pitrou@port.ac.uk}
\affiliation{Institute of Cosmology and Gravitation,
Dennis Sciama Building,\\
Burnaby Road, Portsmouth, PO1 3FX, United Kingdom}

\author{Jean-Philippe Uzan}
\email{uzan@iap.fr}
\affiliation{${\mathcal{G}}{\mathbb{R}}
\varepsilon{\mathbb{C}}{\mathcal{O}}$, Institut d'Astrophysique
de Paris, UMR 7095-CNRS, Universit\'e Pierre et Marie
Curie-Paris 6, 98bis boulevard Arago, F-75014 Paris, France,\\
and Department of Mathematics and Applied Mathematics,\\
University of Cape Town, Rondebosch 7701, Cape Town, South Africa}

\date{December 2, 2009}

\begin{abstract}
This article provides a general study of the Hamiltonian
stability and the hyperbolicity of vector field models involving
both a general function of the Faraday tensor and its dual,
$f(F^2,F\tilde F)$, as well as a Proca potential for the vector
field, $V(A^2)$. In particular it is demonstrated that theories
involving only $f(F^2)$ do not satisfy the hyperbolicity
conditions. It is then shown that in this class of models, the
cosmological dynamics always dilutes the vector field. In the
case of a nonminimal coupling to gravity, it is established that
theories involving $R f(A^2)$ or $Rf(F^2)$ are generically
pathologic. To finish, we exhibit a model where the vector field
is not diluted during the cosmological evolution, because of a
nonminimal vector field-curvature coupling which maintains
second-order field equations. The relevance of such models for
cosmology is discussed.
\end{abstract}
\maketitle
\section{Introduction}

Inflation~\cite{inflation1} is usually invoked to explain the
isotropy and homogeneity of our universe. In particular it has
been demonstrated that if the dynamics of the universe during
inflation is dominated by a scalar field, any primordial spatial
anisotropy is washed out, both at the background
level~\cite{bianchinflation} and perturbation
level~\cite{ppu1,ppu2,peloso1}. Several features of the cosmic
microwave background (CMB) temperature anisotropies seem however
not to be fully consistent with this prediction. This
includes~\cite{CMBprob} the low quadrupole (although its
statistical relevance is questionable), the alignment of the
lowest multipoles and an asymmetry in power between the
northern and southern hemispheres.

It has been suggested that this may be related to an early
anisotropic expansion during the inflationary
phase~\cite{peloso0}. In such a case, it can only lead to an
observable anisotropy in the CMB at the largest angular scales at
the price of a fine tuning on the number of e-folds during
inflation~\cite{ppu1,ppu2,peloso1}. A natural extension of such
an anisotropic expansion is to introduce other matter fields,
besides the inflaton, having the property to source the shear.
This is the case of vector fields~\cite{acw,kanno,yoko},
2-forms~\cite{mangano} or axions~\cite{axion}.

However, vector fields are usually diluted by the cosmological
expansion, both during inflation and the matter era. Indeed,
in a Friedmann-Lema\^{\i}tre spacetime, with
metric\footnote{Throughout this paper, we use the sign
conventions of Ref.~\cite{MTW}, notably the mostly-plus
signature.}
\begin{equation}\label{FLmetric}
\mathrm{d} s^2 = -\mathrm{d} t^2
+ a^2(t)\gamma_{ij}\mathrm{d} x^i\mathrm{d} x^j,
\end{equation}
$t$ being the cosmic time, $a$ the scale factor and $\gamma_{ij}$
the comoving spatial metric, the spatial homogeneity implies that
the only nonvanishing component of the Faraday tensor is
$F_{0i}$. The Maxwell equation reduces to $\ddot A_i + H\dot
A_i=0$ [see Sec.~\ref{secCD1} below for a detailed discussion].
Thus $A^2\propto t^{2-4p}$ if the scale factor scales as $t^p$
and $A^2\rightarrow0$ during the matter era ($p=2/3$) and during
inflation ($p>1$).

This well known fact led to the conclusion that in order to
construct inflationary models driven by a vector field, and even
to have a slow-rolling vector field during inflation, one needs
to include either a potential to the vector
field~\cite{ford,lewis,dimo} or a nonminimal
coupling~\cite{golovnev,chiba}. The stability of these models is
actually an ongoing
debate~\cite{pelosoinstable,pelosoinstable2,golovnev2,peloso3,golovnev3}.
Most of these models have been extended to higher
forms~\cite{2form0,2form1,2form2} and also to models of dark
energy~\cite{mota,mota2,picon,boehmer,jim3,jim1}, which are
essentially the same models applied to the late time dynamics of
the cosmological expansion.

Vector fields are thus central ingredients in various
cosmological models for both the inflationary era and the recent
acceleration. Needless to recall that they also play a key r\^ole
in various extensions of general relativity, with the
vector-tensor theories~\cite{hellings,willVect,Willbook} and more
recently the tensor-vector-scalar theory~\cite{teves} that aims
at reproducing the MOND phenomenology, although they have several
theoretical and experimental
difficulties~\cite{Bruneton-Gef,Bruneton}.

The goal of this article is twofold. First, we want to revisit
the dynamics of vector fields during inflation and take the
opportunity to clarify the structure of theories with
nonminimally coupled vector fields. A fundamental theory should
satisfy two necessary conditions: the boundedness by below of its
Hamiltonian\footnote{More precisely, the spatial integral of the
Hamiltonian density over any localized state should be bounded by
below. Since such localized states may be constructed from a
superposition of sinusoids, at least at linear order, one may
also compute the Hamiltonian density for such spatial sinusoids.}
(otherwise the theory is unstable \cite{woodardtalk}), and the
hyperbolicity of the field equations (so that the Cauchy problem
is well posed \cite{choquet}). We will derive below the
implications of these two conditions on the vector-field theories
we will consider. Of course, as soon as these theories are
assumed to be effective ones, then such conditions need to be
satisfied only in their domain of validity, but this is still
quite constraining.

Section~\ref{sec2} starts by analyzing theories with a minimally
coupled vector field and a quadratic kinetic term, allowing for a
Proca potential, and focuses in a second part on nonlinear
functions $f(F^2,F \tilde F)$ of the Faraday tensor and its dual.
We then consider different classes of nonminimally coupled
theories in Sec.~\ref{sec3}. To finish, we emphasize in
Sec.~\ref{sec5} that there still exist models which allow a
vector field to be slow-rolling, hence offering an interesting
cosmological phenomenology.

Before we start, let us stress that our analysis restricts to
cases where the vector field $A_\mu$ is not of constant norm, and
we refer to Ref.~\cite{jacobson} where such a case was
investigated in depth. Let us also stress that the Hamiltonian
analysis is more powerful than a perturbative analysis around a
particular background since the latter can only demonstrate the
local stability or instability. Hence our analysis will
generalize in many ways some recent
results~\cite{pelosoinstable,pelosoinstable2,golovnev2,peloso3,golovnev3}
concerning the stability of vector-field models.

\section{Minimally coupled theories}\label{sec2}

\subsection{Lagrangian and equations of motion}\label{secIIA}

As a starting point, let us consider a minimally coupled vector
field, whose kinetic term is quadratic in its first derivatives,
and including a potential $V(A^2)$, where $A^2 \equiv A_\mu
A^\mu$. The most general kinetic term \textit{a priori} includes
a linear combination of $(\nabla_\mu A_\nu)(\nabla^\mu A^\nu)$,
$(\nabla_\mu A_\nu)(\nabla^\nu A^\mu)$, and $(\nabla_\mu
A^\mu)^2$. However, the last term can be integrated by parts as
\begin{eqnarray}\label{e:kintegr}
\int \mathrm{d}^4x\sqrt{-g} (\nabla_\mu A^\mu)^2&=&
\int \mathrm{d}^4x\sqrt{-g}\bigl[
(\nabla_\mu A_\nu)(\nabla^\nu A^\mu)
\nonumber\\
&&+R^{\mu\nu}A_\mu A_\nu \bigr],
\end{eqnarray}
so that only a linear combination of the first two terms needs to
be considered in flat spacetime. However, in curved spacetime,
the extra term $R^{\mu\nu}A_\mu A_\nu$ is a particular nonminimal
coupling to gravity.

Let us first recall that, in flat spacetime, the only ghost-free
vector theory in the above class is the standard Maxwell
Lagrangian (called Proca Lagrangian in the massive
case~\cite{EulerH2})
\begin{equation}\label{e.proca}
\mathcal{L}_\text{Maxwell}= -\frac{1}{4} F^2,
\end{equation}
where $F^2\equiv F_{\mu\nu}^2$, and $ F_{\mu\nu} =
\partial_{\mu}A_{\nu}-\partial_{\nu}A_{\mu}$ is the Faraday
tensor. Indeed, if we consider a Lagrangian
\begin{equation}\label{e.Lquad}
\mathcal{L} = \alpha (\partial_\mu A_\nu)^2
+ \beta (\partial_\mu A_\nu)(\partial^\nu A^\mu) - V(A^2),
\end{equation}
we deduce from $F_{0i}=\dot A_i -\partial_iA_0$ that the
conjugate momenta $\pi^\mu \equiv \partial \mathcal{L}/\partial
\dot A_\mu$ read
\begin{equation}\label{e.pis}
\pi^0 = 2(\alpha+\beta)\dot A_0,
\qquad
\pi^i= -2\alpha\dot A_i -2\beta\partial_iA_0.
\end{equation}
If $\alpha+\beta \neq 0$, the field $A_0$ is thus dynamical. This
can also be illustrated by writing the Euler-Lagrange equation
deriving from (\ref{e.Lquad})
\begin{equation}\label{e.vs1}
\alpha\Box A^\nu + \beta\partial^\nu\phi = -V' A^\nu,
\end{equation}
together with its divergence
\begin{equation}\label{e.vs2}
(\alpha+\beta)\Box\phi = -\partial_\lambda(V'A^\lambda),
\end{equation}
where $\phi\equiv\partial_\mu A^\mu$ and $V' \equiv \mathrm{d}
V/\mathrm{d}(A^2)$. Although the replacement of the derivative
$\partial_\mu A^\mu$ by a scalar field would be
illicit\footnote{\label{footnote2a}Redefining a derivative as a
fundamental field in a Lagrangian obviously loses some dynamics,
as illustrated by the trivial case of a scalar-field kinetic term
$\mathcal{L} = -(\partial_\mu \varphi)^2$, which would give an
adynamical vector $\mathcal{L} =-V_\mu^2$ if one redefined $V_\mu
\equiv \partial_\mu \varphi$.} in the Lagrangian (\ref{e.Lquad}),
one may do so in the field equations, and
Eqs.~(\ref{e.vs1})--(\ref{e.vs2}) show that the model describes a
transverse vector field ($A_\mu$ with $\partial_\mu A^\mu=0$)
together with a scalar degree of freedom $\phi$. These equations
also underline that some degrees of freedom become nondynamical
when either $\alpha = 0$ or $\alpha+\beta = 0$, as will be
discussed below.

Let us first consider the generic case where $\alpha \neq 0$ and
$\alpha+\beta \neq 0$. Then the Hamiltonian density $ \mathcal{H}
\equiv \pi^\mu \dot A_\mu - \mathcal{L}$ takes the form
\begin{eqnarray}\label{e.Hquad}
\mathcal{H} &=& \frac{(\pi^0)^2}{4(\alpha+\beta)}
-\frac{(\pi^i + 2\beta \partial_i A_0)^2}{4\alpha}
+\alpha (\partial_iA_0)^2
\nonumber\\
&&-(\alpha+\beta)(\partial_iA_j)^2+\frac{\beta}{2}F_{ij}^2
+ V(A^2).
\end{eqnarray}
Since $\pi^i$ and $\partial_i A_0$ are independent from each
other, and since the quadratic form $-x^2/\alpha + \alpha y^2$ is
not positive definitive (whatever the sign of $\alpha$), we
conclude that $\mathcal{H}$ can take arbitrary large and negative
values, and thereby that the theory is unstable. This is the
well-known result that the massive vector $A_\mu$ contains three
modes of positive energy, but also an extra helicity-0 ghost.

On the other hand, if $\alpha+\beta = 0$, which corresponds to
the usual Maxwell Lagrangian (\ref{e.proca}), then
Eq.~(\ref{e.pis}) yields the primary constraint $\pi^0 = 0$ and
the scalar mode is no longer dynamical. We will recall in
Eq.~(\ref{e.HProca}) below the standard result that $-\alpha =
\beta > 0$ and $V' \geq 0$ are necessary conditions for the
Hamiltonian to be bounded by below.

The other particular case for which expression (\ref{e.Hquad})
for the Hamiltonian cannot be used is when $\alpha = 0$. After
integration by parts, this corresponds to a simple kinetic term
of the form $\beta (\partial_\mu A^\mu)^2$. The conjugate momenta
read then $\pi^0 = 2\beta (\dot A_0 - \partial_i A^i)$ and $\pi^i
= 0$, so that only the helicity-0 degree of freedom contained in
the vector $A_\mu$ is now dynamical. This case should thus be
considered as a \textit{scalar} theory rather than a vector one
(although it differs from standard scalar theories because of the
secondary constraint $2 A_i V' = \partial_i \pi ^0$ imposed by
the field equations). We will thus not consider it any longer in
this paper. Let us just mention that the first term contributing
to the Hamiltonian density $\mathcal{H} = (\pi^0+2\beta\partial_i
A^i)^2/4\beta - \beta (\partial_i A^i)^2 +V$ can obviously be
made positive by choosing $\beta > 0$, but that this does not
suffice to guarantee the stability of the model because the
second term is then negative. The fact that it is not independent
from $\pi^0$ complicates the analysis, but when $V' =
\text{const.} > 0$, for instance, it is easy to build consistent
initial conditions such that $\mathcal{H} \rightarrow -\infty$,
thereby proving that the model is unstable in such a case.

This analysis underlines that vector-field theories are
generically unstable when their kinetic term does not respect the
gauge invariance $A_\mu\rightarrow A_\mu+\partial_\mu\lambda$,
because the $A_0$ component is then a ghost degree of
freedom\footnote{Let us recall that a \textit{ghost} is defined
as a field with negative kinetic energy, not to be confused with
a \textit{tachyon}, a field with negative mass squared, or more
generally with a potential which is unbounded by below. Both
cases correspond to unstable models, but tachyons involve a time
scale whereas the presence of ghosts implies an instantaneous
disintegration of vacuum in quantum mechanics~\cite{woodardtalk}.
Note that a field may be both a tachyon and a ghost, but that the
corresponding model is then even more unstable.}. Before studying
nonlinear vector actions, let us underline that the above
Hamiltonian analysis is fully changed in the case of a
constant-norm vector field; see Ref.~\cite{jacobson} for a
detailed analysis of this interesting case.

\subsection{Function of $F^2$}\label{secIIB}

Let us thus consider now nonlinear functions of $F^2$, i.e.,
gauge-invariant kinetic terms by construction, in Lagrangians of
the form
\begin{equation}\label{e.fdeproca0}
\mathcal{L}= - f\left(F^2\right) - V(A^2).
\end{equation}
The associated field equation for the vector field is then simply
given by
\begin{equation}\label{e.max1}
\nabla_\mu\left(f' F^{\mu\nu} \right)=\frac{1}{2}V' A^\nu,
\end{equation}
where a prime denotes a derivative with respect to the argument
of the function, namely $f' \equiv \mathrm{d} f/\mathrm{d}(F^2)$
and as before $V' \equiv \mathrm{d} V/\mathrm{d}(A^2)$. Note that
$f'$ should never vanish otherwise the Cauchy problem would be
ill-posed. From the definition of the Faraday tensor, we always
have
\begin{equation}\label{e.partF1}
\partial_\alpha F_{\mu\nu} + \partial_\mu F_{\nu\alpha} +
\partial_\nu F_{\alpha\mu} =0,
\end{equation}
and the divergence of Eq.~(\ref{e.max1}) implies
\begin{equation}
\nabla_\nu (V' A^\nu) = 0.
\end{equation}
When $V'\not=0$, this is an extra constraint that arises from the
fact that the action is no more invariant under $A_\mu\rightarrow
A_\mu + \partial_\mu\lambda$, even if the kinetic term
independently is.

\subsubsection{Hamiltonian analysis}

Since we have $F^2=F_{ij}^2-2F_{0i}^2$ in Minkowski spacetime,
the conjugate momenta read
\begin{equation}\label{e.pi01}
\pi^0 = 0,
\end{equation}
which is a primary constraint, and
\begin{equation}\label{e.pii1}
\pi^i = 4f'(F^2)\times(\dot A_i -\partial_i A_0) = 4 f' F_{0i}.
\end{equation}
Since $\dot A_0$ does not appear in
Lagrangian~(\ref{e.fdeproca0}), $A_0$ is an auxiliary field. This
means that the field equation for $A_0$ involves no time
derivatives and can be used as a constraint that eliminates a
field variable, in the case at hand $A_0$.

The Hamiltonian density is thus given by
\begin{equation}
\mathcal{H} = \frac{\pi_i^2}{4f'}
+ \pi^i\partial_iA_0+f(F^2)+V(A^2).
\end{equation}
The $A_0$ dependency of $\mathcal{H}$ can be eliminated by first
performing an integration by parts (in which $\pi^i\partial_iA_0$
becomes $-A_0\partial_i\pi^i$) and then using the secondary
constraint $[\pi_0,\mathcal{H}] = 0$. This secondary constraint ensures
that the primary constraint (\ref{e.pi01}) is consistent with the
equations of motion, and it takes the form\footnote{Note that the
constraint~(\ref{e.dp1}) will be general for any theory in which
$\pi^0=0$ and
$$
\frac{\partial\mathcal{L}}{\partial \partial_iA_0} =
- \frac{\partial\mathcal{L}}{\partial \dot A_i},
$$
since then the Euler-Lagrange equation implies
$$
\partial_i\pi^i = - \frac{\partial\mathcal{L}}{\partial A_0}.
$$
This is the case of all theories in which the kinetic term of the
vector field involves only functions of $F^2$ and $F\tilde F$
(see Sec.~\ref{secEH} below).\label{footnote2}}
\begin{equation}\label{e.dp1}
\partial_i\pi^i = -2V' A_0.
\end{equation}
Actually, it turns out to be the Euler-Lagrange equation
(\ref{e.max1}) for $\nu = 0$, rewritten in terms of conjugate
momenta, and it reduces to the Gauss law when $V' = 0$. Note
that, in general, there may be further constraints arising from
the consistency of the secondary constraints with the equation of
motion, and so on. The distinction between primary and secondary
is not important and they are just constraints that we consider
on the same footing. It follows that
\begin{equation}\label{e.HfdeF}
\mathcal{H} = \frac{\pi_i^2}{4f'}
+\frac{(\partial_i\pi^i)^2}{2V'}+f(F^2)+V(A^2),
\end{equation}
if we assume that $V'\not=0$. [In the case where $V' = 0$, then
$\partial_i\pi^i = 0$ from Eq.~(\ref{e.dp1}), so that $
\mathcal{H}$ does not involve any term $\propto
(\partial_i\pi^i)^2$.] This is a function of the field $A_i$, its
spatial derivatives $\partial_iA_j$, its conjugate momentum
$\pi^i$ and its derivatives $\partial_i\pi^j$, since the argument
of the function $f$ can be expressed as
$F^2=F_{ij}^2-2\pi_i^2/(4f')^2$ and $A_0$ can be eliminated by
resolving Eq.~(\ref{e.dp1}); hence
$\mathcal{H}[A_i,\partial_iA_j,\pi^i,\partial_i\pi^j]$.

Equation (\ref{e.HfdeF}) shows that it is \textit{necessary} that
$f'$ be positive for $\mathcal{H}$ to be bounded by below.
Indeed, if there existed a value, say $\bar F^2$, where $f'(\bar
F^2) <0$, then one could construct initial conditions where
$\pi_i^2 \rightarrow \infty$ and $F_{ij}^2 \rightarrow\infty$
while keeping $\bar F^2=F_{ij}^2-2\pi_i^2/(4f')^2$ constant. The
first term of the r.h.s. of Eq.~(\ref{e.HfdeF}) would then tend
towards $-\infty$ whereas the other ones would remain finite.

Similarly, $V'$ \textit{must} also be positive for $\mathcal{H}$
to be bounded by below. Indeed, using the secondary constraint
(\ref{e.dp1}), the contribution of the potential to
Eq.~(\ref{e.HfdeF}) reads $(\partial_i\pi^i)^2/2V' + V = 2 A_0^2
V'(A^2) + V(A^2)$. If there existed a value, say $\bar A^2$,
where $V'(\bar A^2) <0$, then one could choose initial conditions
where $A_0^2 \rightarrow \infty$ and $A_i^2 \rightarrow\infty$
while keeping $\bar A^2=A_i^2-A_0^2$ constant, and the
Hamiltonian would thus diverge towards $-\infty$.

On the other hand, note that the potential $V$ itself does not
\textit{need} to be bounded by below, contrary to what one may
naively believe from Eq.~(\ref{e.HfdeF}). Indeed, the positive
contribution $2 A_0^2 V'(A^2)$ can compensate negative ones
coming from $V(A^2)$. For instance, for a monomial $V(A^2) = k
(A^2)^n$, where $k$ and $n$ are constants, the contribution of
the potential to the Hamiltonian reads $k \left[(2n-1) A_0^2 +
A_i^2\right] (A^2)^{n-1}$, therefore it is bounded by below if $k
\geq 0$ and $n$ is a positive odd integer. In such a case, $V' =
k n (A^2)^{n-1}$ is consistently positive, but not $V$ itself
since it can have any sign. The particular case $n=1$ corresponds
to the standard massive Proca field, with $V = \frac{1}{2}m^2
A^2$, i.e., $2V' = m^2 > 0$. Then $V = -\frac{1}{2}(\partial_i
\pi^i)^2/m^2 + \frac{1}{2}m^2 A_i^2$ contains a negative term
which can blow up for some specific initial conditions, but it is
counterbalanced by the second term of (\ref{e.HfdeF}),
$+(\partial_i \pi^i)^2/m^2$. The above example of a monomial also
illustrates that $V' \geq 0$ is not a \textit{sufficient}
condition. Indeed, if one chose $k < 0$ and $n$ odd and negative,
then $V'$ would always be positive but $\mathcal{H}$ would
diverge towards $-\infty$ for initial conditions such that
$\partial_i \pi^i = 0$ and $A_i^2 \rightarrow \infty$.

Some negative contributions coming from $f(F^2)$ may also be
compensated by $\pi_i^2/4f' $. This is again what happens in the
massive Proca (or pure electromagnetic) case, where $f(F^2) =
F^2/4=F_{ij}^2 -\pi_i^2/2$ but $\pi_i^2/4f' = \pi_i^2$, so that
\begin{equation}\label{e.HProca}
\mathcal{H} = \frac{1}{2}\pi_i^2
+ \frac{(\partial_i\pi_i)^2}{2 m^2}+\frac{1}{4}F_{ij}^2
+ \frac{1}{2}m^2 A_i^2
\end{equation}
is clearly positive.

Since there is no obvious necessary \textit{and sufficient}
conditions warranting that the Hamiltonian (\ref{e.HfdeF}) is
bounded by below in the most general case, this should be checked
explicitly for any specific theory at hand, recalling that
compensations between terms often occur.

\subsubsection{Hyperbolicity}\label{hyperbolicity}

The second necessary condition that a field theory
(\ref{e.fdeproca0}) should satisfy, is that its field equations
(\ref{e.max1}) are hyperbolic, i.e., that their second
derivatives are of the form $G^{\mu\nu}\partial_\mu\partial_\nu$,
with $G^{\mu\nu}$ an effective metric of signature $-+++$ (its
timelike direction, corresponding to the negative eigenvalue,
should also be consistent with the standard time direction of
$g^{\mu\nu}$). These second derivatives can be written as an
operator acting on the vector field $A_\mu$,
\begin{equation}\label{e.OpToDiagonalize}
\left[f'\times\left(\delta^\nu_\sigma\Box
-\partial_\sigma\partial^\nu\right)
+ 4 f'' F^{\mu\nu}F^\rho_{\hphantom{\rho}\sigma}
\partial_\mu\partial_\rho\right]A^\sigma.
\end{equation}
Our first difficulty, with respect to the better studied case of
scalar ``k-essence'' Lagrangians~\cite{Bruneton-Gef,kess1,kess2}
is that $A_\mu$ has four components and that the above operator
is not diagonal. In order to diagonalize it, it is convenient to
first remove the $-f'\partial_\sigma\partial^\nu$ contribution to
Eq.~(\ref{e.OpToDiagonalize}) by fixing the Lorenz gauge, namely
by adding $\lambda (\partial_\mu A^\mu)^2$ to Lagrangian
(\ref{e.fdeproca0}), where $\lambda$ is a Lagrange multiplier. In
this gauge, the operator (\ref{e.OpToDiagonalize}) becomes of the
form $f'\Box\mathbbm{1} +4f''|v\rangle\langle v|$, where the
Dirac ket $|v\rangle$ represents $F^{\mu\nu}\partial_\nu$. One
finds thus immediately that its four eigenvalues (still as an
operator) are three times $f'\Box$, and once $f'\Box +4f''\langle
v|v\rangle = f'\Box + 4 f''
F^{\mu\rho}F^\nu_{\hphantom{\nu}\rho}\partial_\mu\partial_\nu$,
this fourth ``eigenoperator'' acting in the direction of
$|v\rangle$. Obviously, the operator $f'\Box$ is hyperbolic of
signature $-+++$ if and only if
\begin{equation}\label{e.fpPositive}
f' > 0.
\end{equation}
The fourth eigenoperator may be written as
$G^{\mu\nu}\partial_\mu\partial_\nu$, where $G^{\mu\nu}\equiv f'
g^{\mu\nu}+ 4 f'' F^{\mu\rho}F^\nu_{\hphantom{\nu}\rho}$ is an
effective metric in which the fourth component of the vector
$A_\mu$ (in our specific diagonalizing basis) propagates. The
simplest way to analyze its hyperbolicity, and to ensure that its
timelike direction is consistent with the one of $g^{\mu\nu}$, is
to diagonalize the matrix $G^{\mu\rho}g_{\rho\nu}$ and impose
that its four eigenvalues are positive. [Note that our analysis
uses two different diagonalizations: first a $4\times 4$
\textit{matrix}, with operator values, acting on the vector
$A_\mu$; now the quadratic differential \textit{operator}
$G^{\mu\nu}\partial_\mu\partial_\nu$, acting on one particular
component of $A_\mu$. It happens that $G^{\mu\rho}g_{\rho\nu}$ is
again a $4\times 4$ diagonalizable \textit{matrix}.] These
eigenvalues read
\begin{equation}\label{e.racine}
f' + f'' F_{\mu\nu}^2\pm f''\sqrt{\left(F_{\mu\nu}^2\right)^2
+\left(F_{\mu\nu}\tilde F^{\mu\nu}\right)^2},
\end{equation}
where $\tilde F_{\mu\nu}$ is the dual of the Faraday tensor,
\begin{equation}\label{e.defFtilde}
\tilde F_{\mu\nu} = \frac{1}{2}
\varepsilon_{\mu\nu\rho\sigma}F^{\rho\sigma},
\end{equation}
$\varepsilon_{\mu\nu\rho\sigma}$ being the totally antisymmetric
Levi-Civita tensor such that $\varepsilon_{0123} = +1$. An
elegant way to derive these eigenvalues is to separate
$F_{\mu\nu}$ into standard electric ($E^\mu$) and magnetic
($B^\mu$) field contributions according to an observer with unit
velocity $u^\mu$. Then, in the generic case where $\mathbf{E}$
and $\mathbf{B}$ are not parallel, one may study the action of
the operator
$F^{\mu}_{\hphantom{\mu}\rho}F^{\rho}_{\hphantom{\rho}\sigma}$ on
the four linearly independent vectors $E^\mu$, $B^\mu$, $u^\mu$
and $g^{\beta\mu}\epsilon_{\alpha\beta\gamma\delta}u^\alpha
E^\gamma B^\delta$, and one finds that the spaces spanned by the
first two and the last two are stable under the action of this
operator. In other words, its matrix is constituted of two
$2\times2$ blocks. Its eigenvalues are then easy to compute, and
they happen to be the same for each block. [The particular cases
where $\mathbf{E}$ and $\mathbf{B}$ are parallel or one of them
vanishes are easier to study along the same lines, and one can
check that the result (\ref{e.racine}) remains valid.]

The simultaneous positivity of eigenvalues (\ref{e.racine})
imposes thus the subtle second condition
\begin{equation}\label{e.fsPositive}
F_{\mu\nu}^2 f'' + f' > |f''|\sqrt{\left(F_{\mu\nu}^2\right)^2
+\left(F_{\mu\nu}\tilde F^{\mu\nu}\right)^2}.
\end{equation}
When $F_{\mu\nu}\tilde F^{\mu\nu} = 0$, i.e., when the electric
and magnetic fields are orthogonal, this inequality imposes both
$f' > 0$ [already necessary in Eq.~(\ref{e.fpPositive}) above]
and $2 F_{\mu\nu}^2 f'' + f' > 0$. This should be compared to the
case of scalar k-essence models, whose Lagrangians are functions
$f(s)$ of the standard kinetic term $s \equiv
(\partial_\mu\varphi)^2$. Then the hyperbolicity of the field
equations implies both $f' > 0$ and $2 s f'' + f' >
0$~\cite{Bruneton-Gef,kess1,kess2}.

The fact that the inequality (\ref{e.fsPositive}) depends on two
\textit{independent} relativistic invariants constructed from the
electric and magnetic fields, namely $F_{\mu\nu}^2
=2(\mathbf{B}^2-\mathbf{E}^2)$ and $F_{\mu\nu}\tilde F^{\mu\nu} =
-4\mathbf{E}\cdot\mathbf{B}$, underlines that it should always be
possible to violate it by choosing appropriate initial conditions
on a Cauchy surface. For instance, if $f''(0) \neq 0$, then one
may choose a configuration where $F_{01} = -F_{10} = F_{23} =
-F_{32}$ and all other components vanish. Then $F^2$ vanishes
whereas $\mathbf{E}\cdot\mathbf{B}$ can be chosen as large as one
wishes. This suffices to violate inequality (\ref{e.fsPositive}),
and thereby to prove that the field equations cannot remain
hyperbolic in all physical situations. If the considered theory
is such that $f''(0) = 0$, i.e., that its Lagrangian does not
contain any term proportional to $(F_{\mu\nu}^2)^2$, then we need
to refine slightly the reasoning: We choose a value of $F^2$ such
that $f(F^2) \neq 0$ and we add to it a contribution
$\mathbf{E}\cdot\mathbf{B}$ increasing the value of the square
root in (\ref{e.fsPositive}), while keeping $F^2 =2
\mathbf{B}^2-\mathbf{E}^2$ constant. The only possibility to
always satisfy inequality (\ref{e.fsPositive}) would be to assume
that $f''(F^2) = 0$ for any $F^2$, so that $f(F^2) = k F^2 +
2\Lambda$ (where $k$ and $\Lambda$ are constants) would merely
describe standard Maxwell (or Proca) theory plus a cosmological
constant.

In conclusion, although theories (\ref{e.fdeproca}) can have a
Hamiltonian (\ref{e.HfdeF}) bounded by below for specific
functions $f(F^2)$, there \textit{always} exist situations in
which the field equations are not hyperbolic, because inequality
(\ref{e.fsPositive}) is violated. The only safe case is the
standard Maxwell Lagrangian (with an optional Proca potential).
Of course, if such models are considered as \textit{effective}
theories, then all the above conditions must be satisfied only in
their domain of validity. But if one uses such an effective
theory in situations where Eq.~(\ref{e.fsPositive}) may be
violated, then it just loses any meaning, since the Cauchy
problem is no longer well-posed.

\subsubsection{Cosmological dynamics} \label{secCD1}

Let us investigate the cosmology of the models described by
Lagrangian~(\ref{e.fdeproca0}). The stress-energy tensor of such
a vector field is given by
\begin{equation}
T_{\mu\nu} = 4f' {F^\lambda}_\mu F_{\lambda\nu}
+2V' A_\mu A_\nu -(f+V) g_{\mu\nu}.
\end{equation}
Different roads can then be followed. In particular, it is clear
that the vector field induces the existence of a particular
spatial direction, in contradiction with the hypothesis of
isotropy underneath the form~(\ref{FLmetric}) of the metric. One
should then consider anisotropic cosmological spacetimes, such as
Bianchi universes, which characterize the anisotropy, or try to
recover isotropy by invoking the existence of $N$ vector fields
with random directions and similar initial
magnitude~\cite{golovnev}.

For the sake of simplicity, we investigate the dynamics of a
{\em test} vector field, the dynamics of which is described by
Lagrangian~(\ref{e.fdeproca0}) in a cosmological spacetime with
metric~(\ref{FLmetric}). We can then always decompose the vector
field as
\begin{equation}
A_\mu =(A_0, aB_i), \qquad
A^\mu =\left(-A_0,\frac{1}{a}B^i\right),
\label{eq:AinTermsOfB}
\end{equation}
with $B^i = \gamma^{ij}B_j$. In Cartesian coordinates,
homogeneity implies that $\partial_i A_\mu =0$ so that the only
nonvanishing component of the Faraday tensor is
\begin{equation}
F_{0i} = \dot A_i = a(\dot B_i + H B_i) \equiv aC_i,
\end{equation}
where $H \equiv \dot a/a$ denotes the Hubble function. As
expected, $A_0$ will not enter the equation of evolution and, as
long as $V'\not = 0$, the field equation~(\ref{e.max1}) implies
in Cartesian coordinates that $A^0=0$ and
\begin{equation}
f' \left(\dot F_{0i} + HF_{0i} \right)
+f'' \partial_0(F^2) F_{0i} = -\frac{1}{2}V' A_i.
\end{equation}
Since $F^2 = 2F^{0i}F_{0i} = - 2 F_{0j}F_{0k}\gamma^{jk}/a^2 = -2
C_iC^i=-2C^2$, this equation rewrites as an equation for $B_i$ as
\begin{eqnarray}\label{e.evoBi}
&&\ddot B_i
+ \left[ 3H - 2\frac{f''}{f'}\partial_t(C^2)\right]\dot B_i
\nonumber\\
&&+ \left[ (2H^2+\dot H) + \frac{V'}{2f'}
- 2 H \frac{f''}{f'}\partial_t(C^2)\right] B_i = 0,\quad
\end{eqnarray}
where we use that $f'$ should not vanish, or equivalently as the
system
\begin{eqnarray}
&& \dot C_i + 2\left[H
- \frac{f''}{f'}\partial_t(C^2)\right]C_i
= -\frac{1}{2}\frac{V'}{f'}B_i
\label{e.syst1} \\
&& \dot B_i + HB_i = C_i. \label{e.syst2}
\end{eqnarray}
In that particular case, we deduce that the energy density of the
field, $\rho_A=-T^0_0$, is
\begin{equation}\label{e.rhoA}
\rho_A = 4 f' C_i^2 + f + V.
\end{equation}
Note that the isotropic pressure $P_A=T^i_i/3$ is given by
\begin{equation}\label{e.PA}
P_A = -\frac{4}{3} f' C_i^2 + \frac{2}{3}V'B_i^2- f - V.
\end{equation}
For such a vector field the pressure is however not isotropic and
there is a contribution of the vector field to the anisotropic
stress (i.e. the transverse and traceless part of the
stress-energy tensor)
\begin{equation}
\pi^i_j = - 4 f' \left(C_iC^i
- \frac{1}{3}C^2\delta^i_j \right)
+ 2V'\left(B^iB_j - \frac{1}{3}B^2\delta^i_j\right).
\end{equation}
{}From the expression of the energy density and anisotropic
stress, we see that, in order for the vector field to play any
significant r\^ole, one needs either $C_i$ or $B_i$ not to be
diluted during the expansion\footnote{In particular, if one
relaxes the hypothesis of isotropy and describes the universe by
a Bianchi~I space-time with metric
$$
\mathrm{d} s^2 = -\mathrm{d} t^2
+ a^2(t)\gamma_{ij}(t)\mathrm{d} x^i \mathrm{d} x^j,
$$
the shear $\sigma_{ij}=\frac{1}{2}\dot\gamma_{ij}$ is sourced by
this anisotropic stress and evolves as $\dot\sigma^i_j +
3H\sigma^i_j = 8\pi G \pi^i_j$ so that a nonvanishing anisotropic
stress can source the shear which decays as $a^{-3}$ otherwise;
see Ref.~\cite{ppu1}. Such a vector field, even if it does not
influence the dynamics of inflation can be at the origin of an
homogeneous shear, along the line of Ref.~\cite{kanno}. Note
that the evolution of the vector field is modified so that
equation~(\ref{e.evoBi}) has now a r.h.s. $2\sigma^j_iC_j$,
and $\partial_t(C^2)$ now contains a shear-dependent
contribution $-2 \sigma^{ij} C_i C_j$.}.

In the standard case of the Maxwell theory ($f'=1/4$ and $V=0$),
it is obvious that Eq.~(\ref{e.syst1}) implies that $C_i \propto
a^{-2}$. We then conclude that $\rho_A\propto a^{-4}$ and the
vector field energy density is diluted with respect to the matter
fields driving the expansion of the universe. Indeed, this could
have been deduced from Eqs.~(\ref{e.rhoA})--(\ref{e.PA}) which
imply that, as expected, the equation of state of the homogenous
fluid is $1/3$.

Again, in the Proca case ($f'=1/4$ and $V'\neq0$), the
vector field can play a r\^ole if it is not diluted, i.e., if
$B_i\sim$ const. is a solution of Eq.~(\ref{e.evoBi}). This
happens if the coefficient of $B_i$ is small compared to $H^2$,
and the energy density $\rho_A$ of the vector field is then
almost constant. However this requires that $V'<0$, as
initially proposed in Ref.~\cite{ford}, in contradiction with
the Hamiltonian analysis above.

In the general case, assuming that inflation is described by a de
Sitter phase, i.e., $H$ is constant, the solution $B_i$ constant
can only been reached under the condition that $2H^2 + V'/2f' - 2
H (f''/f')\partial_t(C_i^2)\ll H^2$. This is actually impossible
since $C_i=HB_i$ is also constant and $V'/f'$ is positive. This
can be generalized to the case of slow-roll inflation for which
$\dot H = -\varepsilon H^2$. A configuration with $B_i$ constant
can be reached if
\begin{equation}
2H^2 + \frac{V'}{2f'} + 4\varepsilon H^4 B^2 \frac{f''}{f'}\sim0.
\end{equation}
Since $V'/f'\geq0$, this is possible only if $f''/f'$ is of order
$1/\varepsilon$ and $\varepsilon<0$. Such a fine tuning is very
unnatural since $f$ enters the vector field sector, while
$\varepsilon$ is set by the matter driving the inflationary era.
On the other hand, a configuration with $C_i$ constant requires,
from Eq.~(\ref{e.syst2}), that $aB_i=C_i\int a \mathrm{d} t$. But
Eq.~(\ref{e.syst1}) implies that $2Ha=-(V'/2f')\int a\mathrm{d}
t$, which is impossible as long as the universe is expanding. In
conclusion the vector field cannot play a cosmologically relevant
r\^ole.

This is confirmed by a more general argument. Let us introduce
\begin{equation}
\phi = C_i^2, \qquad \psi = B_i^2,\qquad
C_iB^i =\sqrt{ \phi\psi}\mu,
\end{equation}
with $\mu^2\leq1$ and $\phi$ and $\psi$ positive. From the
system~(\ref{e.syst1})--(\ref{e.syst2}), we can extract the
following set of equations describing the relative evolution of
$B_i$ and $C_i$~:
\begin{eqnarray}
\left(1-4\frac{{f}''}{{f}'}\phi\right)\dot \phi
&=& - 4H\phi - \mu\frac{{V}'}{{f}'}\sqrt{\phi\psi},\label{eq:phidot}\\
\dot \psi &=& - 2H\psi + 2\mu \sqrt{\phi\psi},\label{eq:psidot}\\
\dot\mu&=&\left(\frac{{V}'}{2{f}'} \sqrt{\frac{\psi}{\phi}}
- \sqrt{\frac{\phi}{\psi}}\right)(\mu^2-1).\quad\label{eq:mudot}
\end{eqnarray}
In this system, ${V'}$ is a function of $-\psi$ and ${f}''$ and
${f}'$ are functions of $-2\phi$ so that the system has been
written as a dynamical system. Its fixed point, characterized by
$\dot\phi=\dot\psi=0$, must be such that $V'\psi/f'=-4\phi$, in
contradiction with $V'/f' \geq 0$ unless $\phi = 0$ and $V'\psi =
0$. Even if $V'=0$, setting $\dot \psi = 0$ and $\phi = 0$ in
Eq.~(\ref{eq:psidot}) implies $\psi = 0$ as soon as $H\neq 0$.
Therefore the unique fixed point of this dynamical system
corresponds to $\phi = \psi = 0$, i.e., to a strictly vanishing
vector field.

In conclusion, slow-rolling solutions can be constructed at best
via an unnatural fine-tuning, and moreover, these solutions are
not fixed points of the dynamics. We conclude that such vector
fields will be diluted and play no r\^ole in cosmology.

\subsection{Introducing $F\tilde F$}\label{secEH}

Since the contraction $F\tilde F \equiv F_{\mu\nu}\tilde
F^{\mu\nu}$ appeared in the previous hyperbolicity analysis, we
are naturally led to consider an extension of
theory~(\ref{e.fdeproca0}) of the form
\begin{equation}\label{e.fdeproca1}
\mathcal{L}= - f\left(F^2,F\tilde F\right) - V(A^2).
\end{equation}
In the following, we will set
\begin{equation}\label{e.XY}
X\equiv F^2 \quad\hbox{and}\quad
Y\equiv F\tilde F,
\end{equation}
and denote as $f_X$ and $f_Y$ the partial derivatives of $F$ with
respect to $X$ and $Y$, respectively. The field equation deriving
from (\ref{e.fdeproca1}) can thus be written as
\begin{equation}\label{e.MaxFdual}
\nabla_\nu\left(f_X F^{\mu\nu}+f_Y
\tilde F^{\mu\nu}\right)= \frac{1}{2}V' A^\mu.
\end{equation}
Note that in the particular case in which $f=F\tilde F$, this
equation is empty because of the identity $\partial_\mu\tilde
F^{\mu\nu} = 0$ (i.e., $\mathrm{d}^2 A = 0$ in Cartan's
exterior-derivative notation, namely Maxwell's first set of
equations, $F_{[\mu\nu;\rho]}=0$).

An example of such theories, though it is an effective one, is
the Euler-Heisenberg corrections~\cite{EulerH2,EulerH1} to the
Maxwell Lagrangian (\ref{e.proca}), which take into account the
vacuum polarization. It is given by the Lagrangian
\begin{equation}\label{e.EulerHeisenberg}
\mathcal{L}_{\rm EH} = \frac{\alpha^2}{90 m_e^4}\left[
(F_{\mu\nu}F^{\mu\nu})^2+ \frac{7}{4}(F_{\mu\nu}\tilde F^{\mu\nu})^2
\right],
\end{equation}
where $\alpha$ is the fine-structure constant and $m_e$ the mass
of the electron. It is derived formally as the first term of an
expansion when $\alpha^2 \rightarrow 0$, and its domain of
validity is precisely when such nonlinear corrections remain
small with respect to the standard Maxwell theory
(\ref{e.proca}). In this domain of validity, the Hamiltonian
density is positive and the field equations are hyperbolic,
therefore none of the following discussions need to be done. On
the other hand, as soon as a Lagrangian of form
(\ref{e.fdeproca1}) is considered as defining a fundamental
theory, or when one wishes to study its predictions in a domain
where nonlinear effects are significant, then both the stability
and the well-posedness of the Cauchy problem need to be analyzed
carefully.

\subsubsection{Hamiltonian analysis}\label{subsec000}

As in the previous sections, we need to compute the Hamiltonian
density and we restrict to a Minkowski background spacetime. The
two relativistic invariants (\ref{e.XY}) reduce to
\begin{equation}
X=F_{ij}^2-2F_{0i}^2 \quad\hbox{and}\quad
Y= 2\varepsilon^{ijk}F_{0i}F_{jk},
\end{equation}
where we have set $\varepsilon_{ijk}\equiv\varepsilon_{0ijk}$. It
follows that the conjugate momenta take the form
\begin{eqnarray}
\pi^0 &=& 0, \label{E.piFt1}\\
\pi^i &=& 4f_XF_{0i}-2f_Y\varepsilon^{ijk}F_{jk},\label{E.piFt2}
\end{eqnarray}
and the Hamiltonian density reads
\begin{equation}
\mathcal{H}=\frac{\pi_i^2}{4f_X}
- \frac{f_Y}{2f_X}\varepsilon^{ijk}\pi_iF_{jk}
+\pi^i\partial_iA_0+f+V.
\end{equation}
We are here assuming $f_X \neq 0$, and will consider the
particular case of functions of $Y$ alone in Sec.~\ref{secIIC3}
below. The field equation~(\ref{e.MaxFdual}) reduces, as expected
from the comment in footnote~\ref{footnote2}, to
\begin{equation}\label{e.dipi}
\partial_i\pi^i = -2V'A_0.
\end{equation}
Integrating by part the term $\pi^i\partial_iA_0$ and then using
the secondary constraint to eliminate $A_0$, we end up with a
Hamiltonian density
\begin{equation}\label{e.HFt}
\mathcal{H}=\frac{\pi_i^2}{4f_X} +\frac{(\partial_i\pi^i)^2}{2V'}-
\frac{f_Y}{f_X}\varepsilon^{ijk}\pi_i\partial_jA_k
+f+V.
\end{equation}
[This expression assumes that $V'\neq 0$. When it vanishes, the
second term $\propto(\partial_i\pi^i)^2$ merely disappears
because Eq.~(\ref{e.dipi}) implies $\partial_i\pi^i = 0$.]

As for the simpler case of functions $f(X)$ considered in
Sec.~\ref{secIIB}, one needs to check that the Hamiltonian
density (\ref{e.HFt}) is bounded by below for each specific model
one is considering.

A \textit{necessary} condition is that $f_X$ be positive. Indeed,
in terms of $E_i=F_{0i}$ and $B^i=
\varepsilon^{ijk}\partial_jA_k$, the Hamiltonian may be rewritten
as
\begin{equation}\label{e.Hrewritten}
\mathcal{H}= 4f_X E^2 +\frac{(\partial_i\pi^i)^2}{2V'}
+f-Yf_Y+V.
\end{equation}
Now, one can let $E\rightarrow\infty$ while keeping constant the
arguments $X=2(\mathbf{B}^2-\mathbf{E}^2)$ and $Y= -4{\bf
E}\cdot{\bf B}$ of the function $f$ and its derivatives. [This
can be performed for instance by setting $E=\sqrt{X/2}\sinh p$,
$B=\sqrt{X/2}\cosh p$, $\cos( {\bf E},{\bf B})=-Y/(2X\sinh p\cosh
p)$, and letting the parameter $p\rightarrow\infty$.] Therefore
$\mathcal{H}$ could take arbitrary large and negative values if
we had $f_X < 0$.

Specific \textit{sufficient} conditions may also be written to
ensure that $\mathcal{H}$ is bounded by below. For instance, it
would obviously suffice that $f_X \geq 0$, $V' \geq 0$ and both
$f-Yf_Y$ and $V$ are bounded by below. However, this is far from
being necessary, since the positive contribution coming from
$4f_X E^2$ can compensate a negative one due to $f-Yf_Y$, and
that $(\partial_i\pi^i)^2/2V' + V$ may be bounded by below even
if one of the terms can diverge towards $-\infty$. This is what
happens in the standard Proca case discussed in
Eq.~(\ref{e.HProca}) above.

\subsubsection{Hyperbolicity}\label{hyperbolicity2}

Following the same lines as in Sec.~\ref{hyperbolicity},
equation~(\ref{e.MaxFdual}) for the propagation of the scalar
field can be rewritten as an operator acting on $A^\sigma$,
\begin{eqnarray}\label{e.operatorfXY}
&&f_{X}(\delta^\nu_\sigma\Box - \partial_\sigma\partial^\nu)
\nonumber\\
&&+4\left(f_{XX}F^{\mu\nu}F^\rho_{\hphantom{\rho}\sigma}
+ f_{YY}\tilde F^{\mu\nu}
\tilde F^\rho_{\hphantom{\rho}\sigma}\right)\partial_\mu\partial_\rho
\nonumber\\
&&+4f_{XY}\left(F^{\mu\nu}\tilde F^\rho_{\hphantom{\rho}\sigma}+
\tilde F^{\mu\nu}F^\rho_{\hphantom{\rho}\sigma}
\right)\partial_\mu\partial_\rho.
\end{eqnarray}
For specific particular cases, it is possible to diagonalize its
action as independent operators acting on the components of
$A^\sigma$, and their hyperbolicity can then be analyzed as
before by working in the generic basis $E^\mu$, $B^\mu$, $u^\mu$,
$g^{\beta\mu}\epsilon_{\alpha\beta\sigma\nu}u^\alpha B^\nu
E^\sigma$. However, the first diagonalization is quite involved
and we did not derive the most general conditions which must be
satisfied. Moreover, the analysis of necessary or sufficient
conditions on $f(X,Y)$ ensuring hyperbolicity is also a difficult
task. Therefore, we merely conclude that for each specific model,
one should check both the boundedness by below of the Hamiltonian
density (\ref{e.HFt})--(\ref{e.Hrewritten}) and that the matrix
of operators (\ref{e.operatorfXY}) defines hyperbolic equations
for all physical components of the vector $A^\mu$. However, we
shall see in Sec.~\ref{secIIC4} below that this class of models
(\ref{e.fdeproca1}) does not answer the question we are
addressing in the present paper, i.e., that the vector field is
necessarily diluted by the cosmological expansion.

\subsubsection{Particular case of $f(F\tilde F)$}\label{secIIC3}

The above Hamiltonian analysis assumed that $f_X \neq 0$,
therefore it cannot be followed in the special case where
$f(F\tilde F)$ does not depend on $F^2$. In such a case, it is
straightforward to show that the corresponding Hamiltonian
density is bounded by below only if $V' \geq 0$ and $f - Y f_Y$
is itself bounded by below [the discussion concerning the
potential $V$ is the same as below Eq.~(\ref{e.HfdeF})].
However, as in Sec.~\ref{hyperbolicity} above, the analysis
of the hyperbolicity of the field equations suffices to exclude
these models. Indeed, the field equations read
\begin{equation}
2\, \partial_\mu\left(\tilde F^{\mu\nu}f'\right)
= A^{\nu}V',\nonumber
\end{equation}
that is to say
\begin{equation}\label{e.fieldEqFFdual}
2\, \tilde F^{\mu\nu}\partial_\mu f' = A^{\nu}V'.
\end{equation}
This equation already shows that no propagation of perturbations
can be defined through a spacetime hypersurface where the
background value of $F^{\mu\nu}$ happens to vanish. This suffices
to underline that this class of models is pathological. One may
anyway mimic the analysis of Sec.~\ref{hyperbolicity}, and
diagonalize the differential operator acting on $A_\mu$ in
Eq.~(\ref{e.fieldEqFFdual}). One finds that three out of the four
components do not propagate because they have a strictly
vanishing differential operator. The fourth component is
differentiated by the operator
$G^{\mu\nu}\partial_\mu\partial_\nu$, where $G^{\mu\nu}\equiv 4
f''\,\tilde F^{\mu\rho} \tilde F^{\nu}_{\hphantom{\nu}\rho}$
plays the r\^ole of an effective metric in which perturbations
propagate. The same reasoning as in Sec.~\ref{hyperbolicity}
above then shows that the eigenvalues of the matrix
$G^{\mu\rho}g_{\rho\nu}$ cannot all be simultaneously positive,
and therefore that this last differential operator is not
hyperbolic either. Indeed, one would need to satisfy the strict
inequality
\begin{equation}
\left(F_{\mu\nu}\right)^2 f'' >
|f''|\sqrt{\left(F_{\mu\nu}\right)^2
+ \left(\tilde F^{\mu\nu}F_{\mu\nu}\right)^2},
\end{equation}
which is impossible.

\subsubsection{Cosmological dynamics}\label{secIIC4}

In the particular case of an homogeneous space-time, and as
detailed in Sec.~\ref{secCD1}, the only nonvanishing components
of the Faraday tensor are $F_{0i}$ so that only $\tilde F_{ij}$
is nonvanishing and thus, it implies that $F\tilde F=0$.

As a consequence, the field equation~(\ref{e.MaxFdual}) leads to
the same equation as for the case of a function of $X$ alone that
is to Eq.~(\ref{e.evoBi}) with the function $f(X)$ replaced by
$f(X,0)$. The cosmological dynamics remains unchanged and the
conclusions of Sec.~\ref{secCD1} are not affected.

\subsubsection{Conclusions and remarks}
Our analysis shows that $f(F^2)$ models do not satisfy the
hyperbolicity conditions (unless $f'=\text{const.}$), and that
one must then extend them to $f(F^2,F\tilde F)$. This is needed
if the model is considered as a fundamental theory, but also in
the domain of validity of an effective one. As we shall also see
below, an interesting cosmological phenomenology can generically
be obtained only when the nonlinear corrections become comparable
to the lowest-order $F^2$ kinetic term. The hyperbolicity
conditions need thus to be satisfied in such a case, even if the
model is assumed to be effective.

Independently of these conditions, we also showed that the only
fixed point of the cosmological dynamics corresponds to
$A_\mu=0$, so that the vector field is diluted during the
cosmological expansion, and therefore cannot play any significant
cosmological r\^ole.

We could have imagined more complex terms such as
$F_{\mu\nu}F^{\nu\rho}F_\rho^{\hphantom{\rho}\mu}$ or
$F_{\mu\nu}F^{\nu\rho}F_{\rho\sigma}F^{\sigma\nu}$. However, one
can check that the first combination strictly vanishes while the
second can be rewritten as a function of $F^2$ and $F\tilde F$,
so that our analysis above already considered such possibilities.
Let us also point out that terms such as $(\partial_\mu
A_\nu)(\partial^\nu A^\rho)(\partial_\rho A^\mu)$ generically
excite the helicity-0 ghost degree of freedom.

\subsection{Constant norm vector field}

Given the conclusion of the previous analysis it is interesting
to consider similar theories but with the constraint that the
vector field has a constant norm. General study of constant norm
vector fields have been discussed notably in
Ref.~\cite{jacobson}, and they play an important r\^ole for
instance in the construction of MOND-inspired
theories~\cite{teves,Bruneton-Gef}.

We may thus consider Lagrangians of the form
\begin{equation}\label{e.fdeproca}
\mathcal{L}= - f\left(F^2,F\tilde F\right) - V(A^2)
+\lambda(A^2-v),
\end{equation}
where $\lambda$ is a Lagrange multiplier and $v$ a number. The
extremization of the action with respect to $\lambda$ gives the
constraint
\begin{equation}\label{e.normA}
A^2=v,
\end{equation}
and $A_\mu$ is timelike (resp. spacelike) when $v<0$ (resp.
$v>0$). The norm-fixing term does not change the expression of
the conjugate momenta which are still given by
Eqs.~(\ref{E.piFt1})-(\ref{E.piFt2}). The equation of motion gets
an extra term
\begin{equation}
\partial_i\pi^i = -2V'A_0+2\lambda A_0.
\end{equation}
It cannot be used to eliminate $A_0$ from the Hamiltonian density
since it is now used to fix the value of $\lambda$. Instead, we
use Eq.~(\ref{e.normA}) to get
\begin{equation}
A_0 = \sqrt{A_i^2-v}.
\end{equation}
We conclude that the Hamiltonian density simplifies to
\begin{equation}
\mathcal{H}=\frac{\pi_i^2}{4f_X} - \sqrt{A_i^2-v}(\partial_i\pi^i)-
\frac{f_Y}{f_X}\varepsilon^{ijk}\pi_i\partial_jA_k
+f+V,
\end{equation}
where the only difference with the expression~(\ref{e.HFt}) lies
in the second term. This expression should be compared to the
result of Refs.~\cite{Clayton:2001vy,Bruneton-Gef}.

Following the same approach as in Sec.~\ref{subsec000}, we get
\begin{equation}
\mathcal{H}= 4f_X E^2 - \sqrt{{\bf A}^2-v}\,
{\bm\nabla}\cdot{\bm\pi} +f-Yf_Y+V.
\end{equation}
We conclude that whatever the functions $f$ and $V$, this
Hamiltonian density is not bounded by below because one can
always let $X$, $Y$ and $A^2$ constant while letting
${\bm\nabla}\cdot{\bm\pi}$ go to infinity.

It should be underlined that the above conclusion only applies to
kinetic terms of the form (\ref{e.fdeproca}). As shown in
Ref.~\cite{jacobson}, more general kinetic terms for a
constant-norm vector field, of the form $ c_1 (\partial_\mu
A_\nu)^2 + c_2 (\partial_\mu A^\mu)^2 + c_3 (\partial_\mu
A_\nu)(\partial^\nu A^\mu) + c_4 (A^\mu \partial_\mu A_\nu)^2$,
can be consistent for specific ranges of values of the constant
coefficients $c_{1,2,3,4}$, i.e., define stable and well-posed
field theories and even pass solar-system and binary-pulsar tests
of relativistic gravity. The same analysis has not yet been
generalized to nonlinear functions of such kinetic terms, nor to
variable coefficients (depending on some field).

\section{Nonminimal couplings}\label{sec3}

The results of the previous section drive us to consider theories
with a standard kinetic term. This section focuses on models
satisfying this constraints but involving a nonminimal coupling
to gravity. This class of models is of particular interest in
cosmology because it has been argued that when such a coupling
exists the vector can be slow-rolling~\cite{golovnev} and the
stability of this models has been debated with different
conclusions~\cite{pelosoinstable,pelosoinstable2,golovnev2,peloso3,golovnev3}.

We already saw, in Eq.~(\ref{e:kintegr}) above, that nonminimal
vector-metric couplings of the form $R^{\mu\nu}A_\mu A_\nu$ are
generated by a mere integration by parts of a general vector
kinetic term in curved spacetime. Such a term, together with a $R
A^2$ coupling, has been considered in chapter 5.4 of
Ref.~\cite{Willbook}. In the following, we will not study
$R^{\mu\nu}A_\mu A_\nu$, whose mathematical and phenomenological
consequences are similar to those of $R A^2$. However, we will
consider the more general case of nonlinear couplings to a
function of $A^2$ in Sec.~\ref{secIIIA}, and show that the
corresponding models are unstable. We will also consider
couplings to a function of the Faraday tensor $F$ in
Sec.~\ref{secIIIB}, but underline that instabilities are also
generic in such a case.

\subsection{$A^2$ case}\label{secIIIA}

\subsubsection{Jordan frame}

Let us first consider models of the class
\begin{eqnarray}\label{e.nmc_actionJ}
S&=&\int\mathrm{d}^4x\sqrt{-g}\left[\frac{R}{2\kappa}\Psi(A^2)
-\frac{1}{4}F^2 - V(A^2)\right]\nonumber\\
&&\qquad + S_{\rm matter}[\psi_m;g_{\mu\nu}],
\end{eqnarray}
where $\kappa=8\pi G$, $g_{\mu\nu}$ denotes the Jordan frame
metric, and we define $F^2=F_{\mu\nu}F_{\rho\sigma} g^{\mu\rho}
g^{\nu\sigma}$ and $A^2\equiv A_\mu A_\nu g^{\mu\nu}$. $\Psi$
is an arbitrary positive function and the particular case
$\Psi=1+8\pi G\xi A^2$ has been extensively studied in the
literature~\cite{golovnev,kanno}. $G$ is the bare gravitational
constant. It is not the constant that would be measured in a
Cavendish experiment since the vector field is responsible for an
interaction. As in the case of scalar-tensor
theories~\cite{STref}, the Jordan metric is the ``physical
metric" since the matter fields are universally coupled to
$g_{\mu\nu}$. This metric defines the lengths and times actually
measured by laboratory rods and clocks, since they are made of
matter. All experimental data have their usual interpretation in
this frame.

The equation of motions, obtained by variation with respect to
the vector field, is given by
\begin{eqnarray}\label{e.nmc_max}
\nabla_\mu F^{\mu\nu} - \left(2V' - \frac{R}{\kappa}\Psi'
\right)A^\nu =0
\end{eqnarray}
which generalizes the Maxwell equation. As previously, a prime
denotes a derivative with respect to the argument, $V'\equiv
\mathrm{d} V(X)/\mathrm{d} X$. The divergence of this equation
implies that
\begin{eqnarray}
\nabla_\nu\left[\left(2V' - \frac{R}{\kappa}\Psi'
\right)A^\nu\right]=0,
\end{eqnarray}
which is the standard constraint satisfied by a massive Proca
field, in which $-R \Psi(A^2)/2\kappa$ plays the r\^ole of an
extra contribution to the vector's potential $V(A^2)$.

The Einstein and conservation equations, obtained respectively by
varying with respect to the Jordan metric and the matter fields,
yield
\begin{eqnarray}
&&\Psi(A^2)G_{\mu\nu} - \left(\nabla_\mu\nabla_\nu -
g_{\mu\nu}\Box\right)\Psi(A^2) + R\Psi'(A^2)A_\mu A_\nu~\nonumber\\
&&=\kappa\left[F_{\mu\alpha}{F_\nu}^\alpha - \frac14 g_{\mu\nu} F^2+
2 V' A_\mu A_\nu - V g_{\mu\nu} +T_{\mu\nu}^{\rm mat}\right],
\nonumber\\
\label{e.EinsteinNonMin}\\
&&\nabla_\mu T^{\mu\nu}_{\rm mat}=0,\label{e.nmc_mat}
\end{eqnarray}
the second equation being no surprise since the matter fields are
minimally coupled to the Jordan metric.

On the other hand, Eq.~(\ref{e.EinsteinNonMin}) already exhibits
the deadly problem that this class of models presents: Some
gauge-dependent second derivatives of the vector field $A_\mu$
are generated in the left hand side. They come from the $R
\Psi(A^2)$ term in action (\ref{e.nmc_actionJ}), which breaks the
gauge invariance of the vector's kinetic term. Indeed, the scalar
curvature $R$ contains second derivatives of the metric,
therefore, after integration by parts, second derivatives of
$A_\mu$ which cannot be written in terms of the gauge-invariant
Faraday tensor $F_{\mu\nu}$ (nor its dual $\tilde F_{\mu\nu}$).
We thus expect to excite the generic helicity-0 ghost of
non-gauge-invariant vector theories, as in Sec.~\ref{secIIA}
above. We will see below that this will become explicit thanks to
a change of variables, namely by rewriting the same theory in the
so-called Einstein frame. Equation (\ref{e.EinsteinNonMin}) also
illustrates why this ghost is never noticed when studying linear
perturbations, around a background where $A_\mu = 0$. Indeed, the
gauge-dependent second derivatives are acting on a function of
$A^2$, and therefore disappear at linear order in $A_\mu$. This
is actually already manifest in action (\ref{e.nmc_actionJ}),
since the gauge-dependent terms involving derivatives of $A_\mu$
are of the cubic form $A^2 \partial\partial h$ (where $h$ denotes
schematically a perturbation of the metric), and therefore of
quadratic order in the field equations.

\subsubsection{Einstein frame}
The kinetic terms of the spin-1 and spin-2 degrees of freedom are
not diagonalized in action (\ref{e.nmc_actionJ}), as clearly
illustrated by the field equations
(\ref{e.nmc_max})--(\ref{e.EinsteinNonMin}). As for scalar-tensor
theories, the theory is better analyzed in the so-called Einstein
frame, defined by diagonalizing the kinetic terms. This can be
achieved thanks to a conformal rescaling of the metric
\begin{equation}\label{confTransf}
g_{\mu\nu}^* = \Psi(A^2)g_{\mu\nu}.
\end{equation}
For the sake of clarity, we set $A_*^2= g ^{\mu\nu}_*A_\mu A_\nu$
so that
\begin{equation}\label{Astar}
A_*^2 = \frac{A^2}{\Psi(A^2)},
\end{equation}
which is assumed to be invertible as a function $\Phi(A_*^2)
\equiv A^2$. When performing the conformal transformation and
also replacing $A^2$ in terms of $A_*^2$, we obtain that
action~(\ref{e.nmc_actionJ}) can be rewritten as
\begin{eqnarray}\label{e. nmc_actionE}
S&=&\int\mathrm{d}^4x\sqrt{-g_*}\biggl[\frac{1}{2\kappa}R_*
-\frac{3}{4\kappa}Z^2(A_*^2)(\partial_\mu A_*^2)^2
-\frac{1}{4}F_*^2 \nonumber\\
&&- W(A_*^2)\biggr]
+ S_{\rm matter}\left[\psi_m; B(A_*^2)g^*_{\mu\nu}\right],
\end{eqnarray}
where only use of the Einstein metric $g^*_{\mu\nu}$ is made in
all contractions and in defining the Ricci scalar $R_*$. We
notably define as usual $F_{\mu\nu} = \partial_\mu A_\nu
-\partial_\nu A_\mu$ but $F_*^{\mu\nu} = g_*^{\mu\rho}
g_*^{\nu\sigma} F_{\rho \sigma} $ and $A_*^\mu = g_*^{\mu\nu}
A_{\nu}$. The three functions of $A_*^2$ that appear in this
action are given by
\begin{eqnarray}
B(A_*^2) &\equiv& 1/\Psi(A^2),\\
Z(A_*^2) &\equiv& - \frac{\mathrm{d} \ln B}{\mathrm{d} A_*^2} =
\frac{\Psi(A^2) \Psi'(A^2)}{\Psi(A^2) - A^2 \Psi'(A^2)},\label{Zdef}\\
W(A_*^2) &\equiv& V(A^2)/\Psi^2(A^2).
\end{eqnarray}
The kinetic terms of the vector $A_\mu$ and the tensor
$g^*_{\mu\nu}$ are now diagonalized in action (\ref{e.
nmc_actionE}), in a covariant way. This will allow us to consider
the vector sector alone in in Sec.~\ref{IIIA3} below, say in a
freely falling elevator, to analyze its stability.

Let us however underline a subtlety related to vector fields in
curved spacetime, as soon as their kinetic term is not a mere
function of the Faraday tensor $F_{\mu\nu}$ and its dual $\tilde
F_{\mu\nu}$. Indeed, the contribution proportional to $Z^2$ in
action (\ref{e. nmc_actionE}) involves a cross kinetic term of
the form $\partial A \partial g_*$, because the inverse metric
enters the square $A_*^2 = g_*^{\alpha\beta} A_\alpha A_\beta$.
This can be seen either by writing $\partial_\mu A_*^2 = 2
A_*^\alpha \partial_\mu A_\alpha + A_\alpha A_\beta \partial_\mu
g_*^{\alpha \beta}$ in a non-covariant way, or by recalling
the presence of a Christoffel symbol in the covariant form
$\partial_\mu A_*^2 = 2 A_*^\alpha \nabla^*_\mu A_\alpha$.
This is also illustrated by the Einstein equations deriving from
action (\ref{e. nmc_actionE}), which read
\begin{eqnarray}
\label{eq:EinsteinNonMin}
G^*_{\mu\nu} &=& \kappa \left(T^{*{\rm mat}}_{\mu \nu}+
T^{*{\rm EM}}_{\mu \nu}-W g^*_{\mu\nu}\right)
+\frac{3}{2}
Z^2 \partial_\mu A_*^2 \partial_\nu A_*^2\nonumber\\
&&
-3 \left[ZZ'\left(\partial_\alpha
A_*^2\right)^2 + Z^2 \Box^* A_*^2\right]
A_\mu A_\nu
\nonumber\\
&&-\frac{3}{4} Z^2\left(\partial_\alpha
A_*^2\right)^2 g^*_{\mu\nu},
\end{eqnarray}
where $T_{\rm mat}^{*\mu\nu}\equiv (2/\sqrt{-g_*})(\delta
S_{\mathrm{mat}}/\delta g^*_{\mu\nu})$ is the matter
energy-momentum tensor as defined in the Einstein frame. The
presence of second derivatives of the vector field in
Eq.~(\ref{eq:EinsteinNonMin}), in the form of $\Box^* A_*^2$,
underlines that cross kinetic terms were actually still involved
in action (\ref{e. nmc_actionE}). On the other hand, no curvature
tensor enters the Maxwell equations deriving from action (\ref{e.
nmc_actionE}) in the Einstein frame:
\begin{eqnarray}\label{MaxwellinEinsteinframe}
\nabla^*_\mu F_*^{\mu\nu} &=&A_*^\nu \Bigl[2 W'-\frac{3}{\kappa}
ZZ'\left(\partial_\alpha A_*^2 \right)^2 \nonumber\\
&&- \frac{3}{\kappa} Z^2 \Box^* A_*^2
+\Psi^2 \Psi' T^*_{\rm mat}\Bigr],
\end{eqnarray}
where $T^*_{\rm mat} \equiv g^*_{\mu\nu}T_{\rm mat}^{*\mu\nu}$.
It should be noted that the actual energy-momentum tensor
measured by an observer is the Jordan-frame one, defined as
$T_{\rm mat}^{\mu\nu}\equiv (2/\sqrt{-g}) (\delta
S_{\mathrm{mat}}/\delta g_{\mu\nu})$, and its trace as $T_{\rm
mat}\equiv g_{\mu\nu}T_{\rm mat}^{\mu\nu}$. It is related to its
Einstein-frame counterpart in a nontrivial way, because $B(A_*^2)
g^*_{\mu\nu}$ depends on the Einstein metric $g^*_{\mu\nu}$ also
through $A_*^2 = g_*^{\alpha\beta} A_\alpha A_\beta$. One finds
$T_{\rm mat}^{*\mu\nu} = B^3 (T_{\rm mat}^{\mu\nu} - B' A^\mu
A^\nu T_{\rm mat})$, so that the last term within the square
brackets of Eq.~(\ref{MaxwellinEinsteinframe}) may also be
written as $\Psi^2 \Psi' T^*_{\rm mat} = Z\, T_{\rm mat}$.

Although the kinetic terms are covariantly diagonalized in the
Einstein-frame action (\ref{e. nmc_actionE}), one may be worried
by the non-covariant cross term $\partial A \partial g_*$ it
still contains. Indeed, it is well known that such cross terms
may contribute positively to the kinetic energy of a degree of
freedom. The best known example is Brans-Dicke scalar-tensor
theory, defined by the action $S = \int \mathrm{d}^4 x
\sqrt{-g}\left[\Phi R - (\omega/\Phi)
(\partial_\mu\Phi)^2\right]$, where the spin-0 degree of freedom
carries positive energy provided $\omega > -\frac{3}{2}$. For
$-\frac{3}{2}<\omega<0$, one may thus naively think the scalar
field is a ghost, but the cross kinetic term involved in $\Phi R$
(after partial integration) is enough to guarantee the positivity
of energy. To check that the remaining cross kinetic term of
action (\ref{e. nmc_actionE}) actually does not change our
conclusion of Sec.~\ref{IIIA3} below, let us eliminate it in a
non-covariant way. The clearest way to do so will be to start
again from the Jordan-frame action (\ref{e.nmc_actionJ}), and to
consider perturbations around a given background, keeping
covariant expressions with respect to the background metric. Let
us define $g^\text{full}_{\mu\nu} = g_{\mu\nu} + h_{\mu\nu}$ and
$A^\text{full}_\mu = A_\mu + a_\mu$, and expand
(\ref{e.nmc_actionJ}) to second order in the dynamic
perturbations $h_{\mu\nu}$ and $a_\mu$, using the background
metric $g_{\mu\nu}$ to contract indices or define covariant
derivatives. The kinetic terms of these perturbations then read
\begin{eqnarray}\label{kineticPerturb}
&&-\frac{1}{16\kappa}\Psi(A^2)\nabla_\mu h_{\alpha\beta}
\left(2g^{\alpha\gamma}g^{\beta\delta}
-g^{\alpha\beta}g^{\gamma\delta}\right)
\nabla^\mu h_{\gamma\delta}\nonumber\\
&&+\frac{1}{16\kappa}\Psi(A^2)\left(2\nabla_\nu h^\nu_\mu
-\nabla_\mu h\right)^2\nonumber\\
&&-\frac{1}{2\kappa}\Psi'(A^2)(\nabla_\nu
h ^{\mu\nu} - \nabla^\mu h)
(2 A^\rho \nabla_\mu a_\rho
- A^\rho A^\sigma \nabla_\mu h_{\rho\sigma})\nonumber\\
&&-\frac{1}{4}(\nabla_\mu a_\nu -\nabla_\nu a_\mu)^2,
\end{eqnarray}
where $h \equiv g^{\alpha\beta} h_{\alpha\beta}$ is the trace of
the Jordan metric perturbation. The first two terms of
(\ref{kineticPerturb}) are the standard kinetic term of a spin-2
graviton, multiplied by a global factor $\Psi(A^2)$ depending on
the background vector field $A_\mu$, the fourth term is the
standard Maxwell kinetic term, and the third term exhibits the
cross kinetic terms $\nabla h \nabla a$ generated by the
nonminimal coupling $R \Psi(A^2)$ of action
(\ref{e.nmc_actionJ}). Before diagonalizing these kinetic terms,
let us recall that the general coordinate-invariance of action
(\ref{e.nmc_actionJ}) implies the gauge-invariance of the Jordan
metric perturbation $h_{\alpha\beta}$ (although the Jordan metric
$g_{\alpha\beta}$ does not describe a pure spin-2 degree of
freedom). We may thus fix the harmonic gauge in
Eq.~(\ref{kineticPerturb}) by imposing
\begin{equation}\label{harmonicGauge}
2\nabla_\nu h^\nu_\mu = \nabla_\mu h.
\end{equation}
This choice not only removes the second term of
(\ref{kineticPerturb}), but also simplifies the third term as
\begin{equation}
\frac{\Psi'(A^2)}{4\kappa} \nabla^\mu h\,
(2 A^\rho \nabla_\mu a_\rho
- A^\rho A^\sigma \nabla_\mu h_{\rho\sigma}).
\end{equation}
It is now straightforward to check that the redefinition
\begin{eqnarray}
h^{\text{new}}_{\alpha\beta} &\equiv& h_{\alpha\beta}
+\frac{2\Psi' A^\rho a_\rho}{(\Psi - A^2 \Psi')^2
+ 2 (A^2 \Psi')^2}\nonumber\\
&&\times\left[\left(\Psi + A^2 \Psi'\right)g_{\alpha\beta}
- 4 \Psi' A_\alpha A_\beta\right]
\label{hRedefinition}
\end{eqnarray}
then suffices to eliminate all cross terms $\nabla h^{\text{new}}
\nabla a$. This change of variable differs in several ways from
the conformal transformation (\ref{confTransf}) used above in the
covariant calculation. Indeed, it now contains a ``disformal''
(i.e., non-conformal) contribution proportional to $A_\alpha
A_\beta$. Moreover, it clearly breaks general covariance since
the modification of $h_{\alpha\beta}$ is proportional to the mere
contraction $A^\rho a_\rho$, whereas the expansion of
$A_\text{full}^2 = A^2 + 2 A^\rho a_\rho - A^\rho A^\sigma
h_{\rho \sigma} +\mathcal{O}\left((h,a)^2\right)$ also involves
the projection of the metric perturbation along the background
vector field, $A^\rho A^\sigma h_{\rho \sigma}$. This is one of
the reasons why (\ref{hRedefinition}) allows us to cancel the
cross kinetic term $ \partial A\partial g_*$ we had in the
covariant action (\ref{e. nmc_actionE}). Finally,
Eq.~(\ref{Astar}) happened not to be invertible for the simple
case of $\Psi(A^2) = A^2$, as also illustrated by the vanishing
denominator of definition (\ref{Zdef}), whereas
Eq.~(\ref{hRedefinition}) is always invertible as soon as
$\Psi(A^2)\neq 0$.

It should be noted that the gauge fixing (\ref{harmonicGauge})
is non trivial in terms of the new variable
$h^{\text{new}}_{\alpha\beta}$, since it now also involves the
vector perturbation $a_\rho$. However, as already underlined
above, the general covariance of the Jordan-frame action
(\ref{e.nmc_actionJ}) anyway guarantees this choice is allowed.
It underlines that some cross-kinetic terms (between
$h^{\text{new}}$ and $a$) are actually pure gauge, and cannot
contribute to any physical observable. Replacing now definition
(\ref{hRedefinition}) in (\ref{kineticPerturb}), still in the
gauge (\ref{harmonicGauge}), we can read off the full kinetic
term of the vector perturbation:
\begin{equation}\label{kineticPerturbVector}
-\frac{1}{4}(\nabla_\mu a_\nu -\nabla_\nu a_\mu)^2
-\frac{2}{\kappa}\,
\frac{\Psi \Psi'^2\,(A^\rho\nabla_\mu a_\rho)^2}{(\Psi
- A^2 \Psi')^2 + 2 (A^2 \Psi')^2}.
\end{equation}
This is similar to the expression (\ref{e. nmc_actionE}) we found
in the fully covariant case, with the minor difference of a
global factor $1/\Psi$ for the second term [coming from the fact
that we use the Einstein metric (\ref{confTransf}) to contract
all indices in (\ref{e. nmc_actionE}), whereas we kept the
original Jordan metric $g_{\mu\nu}$ as our present background],
the important difference that the denominator of this second term
contains a contribution $+2 (A^2 \Psi')^2$ in addition to the
square $(\Psi - A^2 \Psi')^2$ coming from $Z^2$ [this change also
comes with a modification of the global numerical factor from $3$
to $2$], and the crucial difference that all cross kinetic terms
have been cancelled. When considering (\ref{e. nmc_actionE}) in a
flat background $g^*_{\mu\nu} = \eta_{\mu\nu}$ (or in a Fermi
coordinate system), we thus get an expression of the same form as
expansion (\ref{kineticPerturbVector}), the only difference being
the precise definition of $Z$. In Sec.~\ref{IIIA3} below, the
nonvanishing of this function $Z$ will be the only needed
information, therefore one may work with the covariant action
(\ref{e. nmc_actionE}) although its kinetic terms are not fully
diagonalized.

Both (\ref{e. nmc_actionE}) and (\ref{kineticPerturbVector}) show
that the mode of $a_\rho$ which is polarized in the direction of
the background $A_\rho$ behaves as if it were a positive-energy
scalar field (see also Sec.~5 of the recent
Ref.~\cite{golovnev3}). However, it is coupled to the other
vectorial modes via the standard Maxwell kinetic term, and we
will see now that this causes a deadly instability of the model.

\subsubsection{Hamiltonian analysis}\label{IIIA3}
The stability analysis of any model is much more easily performed
in the Einstein frame, where the spin 2 and the other degrees of
freedom decouple. As discussed in the previous section, there
still exists a cross kinetic term $\partial A \partial g_*$ in
the covariant action (\ref{e. nmc_actionE}), but eliminating it
in a non-covariant way, as in Eq.~(\ref{kineticPerturbVector}),
keeps the same general form for the vector's kinetic term. Let us
thus consider an action of the form (\ref{e. nmc_actionE}), with
$Z\neq 0$ but maybe different from (\ref{Zdef}), and focus on the
vector's dynamics in a flat geometry $g_{\mu\nu}^*=
\eta_{\mu\nu}$. The conjugate momenta are then given by
\begin{eqnarray}
\pi^0&=&-\frac{3}{\kappa} Z^2 A_0\,
\partial_t(A^2_*),\label{e.pi0nonmin}\\
\pi^i&=&\dot A_i-\partial_i A_0
+\frac{3}{\kappa} Z^2 A_i\,\partial_t(A^2_*).
\end{eqnarray}
Note that at linear order in the field equations (i.e., quadratic
order in the action or the Hamiltonian), we recover $\pi^0 = 0$
as in gauge-invariant vector theories. Therefore the ghost
instability present in the nonminimally coupled models
(\ref{e.nmc_actionJ}) or (\ref{e. nmc_actionE}) cannot be
noticed when studying first-order perturbations (around a
vanishing-vector background).

We deduce that the Hamiltonian density takes the form
\begin{eqnarray}
\mathcal{H}&=&\frac{1}{4}F_{ij}^2
+W+\frac{3}{4\kappa}Z^2\times
\left[\partial_i \left(A_*^2\right) \right]^2
+\frac{\kappa}{12}\left( \frac{\pi^0}{Z A_0}
\right)^2\nonumber\\
&&+\frac{1}{2}\left(\pi^*_{i}
+\frac{\pi^0}{A_0}A_i
+\partial_i A_0 \right)^2
-\frac{1}{2}\left(\partial_i A_0\right)^2\,.
\end{eqnarray}
Since $\pi^0$ is not identically zero in Eq.~(\ref{e.pi0nonmin}),
the $A_0$ component is dynamical and independent from the spatial
components $A_i$. We may thus consider a particular background
such that $A_0 \neq 0$ while $\pi^0=0$, $A_i = 0$ and
$\pi^i=-\partial_i A_0$, and the Hamiltonian density then reads
\begin{equation}
\mathcal{H} \simeq W+\left(\frac{3}{\kappa}
Z^2 A_0^2-\frac{1}{2}\right)\left(\partial_i A_0\right)^2.
\end{equation}
Initial data of the form $A_0 = \varepsilon
\sin(x/\varepsilon^2)$, with $\varepsilon \rightarrow 0$, would
thus make this Hamiltonian density tend towards $-\infty$. This
suffices to show that the nonminimally coupled vector model
(\ref{e.nmc_actionJ}) or (\ref{e. nmc_actionE}) is unstable.

\subsection{$F^2$ case}\label{secIIIB}

We can also consider theories in which the Faraday tensor is
nonminimally coupled to the Ricci scalar. In the Jordan frame,
such theories will have an action of the form
\begin{eqnarray}\label{e.nonminF2}
S&=&\int\mathrm{d}^4x\sqrt{-g}\left[\frac{R}{2\kappa}\Xi(F^2)
- f(F^2)-V(A^2)\right]\nonumber\\
&& + \, S_{\rm matter}[\psi_m;g_{\mu\nu}]
\end{eqnarray}
with the same definitions as in the previous sections.

One may be tempted to introduce the analogue of an Einstein
metric by defining
\begin{equation}
g_{\mu\nu}^* = \Xi(F^2)g_{\mu\nu},
\end{equation}
but since this definition involves \textit{derivatives} of the
vector field, it cannot be used consistently in a Lagrangian (see
footnote \ref{footnote2a} above).

Actually, because the scalar curvature $R$ involves second
derivatives of the metric tensor $g_{\mu\nu}$, action
(\ref{e.nonminF2}) generates \textit{third} derivatives of the
vector field in the metric field equation, and \textit{third}
derivatives of the metric (i.e., covariant derivatives of the
curvature tensor) in the vector field equation. Initial data on a
Cauchy surface should thus contain more information than the
values of the fields and their time derivatives. Therefore, this
class of models must involve some extra degrees of freedom, in
addition to the vector and the metric we wished to introduce.
Such higher derivatives are known to produce generically ghost
degrees of freedom, i.e., to cause the theory to be unstable.
This is notably the consequence of a theorem by
Ostrogradski~\cite{ostro}, well discussed in
Ref.~\cite{woodardtalk}. However, this theorem can be applied
only on so-called ``nondegenerate'' Lagrangians, which produce
\textit{fourth}-order field equations. Therefore, we are here in
a typical case where we expect a serious instability to manifest,
but where we cannot use the generic theorem which proves so
without any ambiguity.

To understand intuitively the instability of a theory defined by
action (\ref{e.nonminF2}), one may consider a toy model involving
two coupled scalar fields in flat spacetime, $\mathcal{L} =
-(\partial_\mu \varphi)^2 -(\partial_\mu \psi)^2 +
\lambda(\partial_\mu \varphi)^2(\partial_\nu \psi)^2$, where
$\lambda$ is a coupling constant (see Sec. V~A of
Ref.~\cite{Bruneton-Gef}). Here $\varphi$ and $\psi$ play the
r\^oles of the metric tensor and of the vector field of
Eq.~(\ref{e.nonminF2}). The corresponding Hamiltonian density
reads $\mathcal{H} = \dot\varphi^2 + \dot\psi^2
+(\partial_i\varphi)^2+(\partial_i\psi)^2 +
4\lambda\dot\varphi^2\dot\psi^2 -\lambda[\dot\varphi^2 +
(\partial_i\varphi)^2] [\dot\psi^2 + (\partial_j\psi)^2]$, and it
can be made arbitrary large and negative whatever the sign of
$\lambda$. Indeed, if $\lambda < 0$, it suffices to choose a
homogeneous configuration $\partial_i\varphi = \partial_i\psi =
0$ and large enough values of $\dot\varphi^2$ and $\dot\psi^2$.
On the other hand, if $\lambda > 0$, instantaneously constant
fields $\dot\varphi = \dot\psi = 0$ with large enough spatial
derivatives $(\partial_i\varphi)^2$ and $(\partial_i\psi)^2$
suffice to make $\mathcal{H}$ tend towards $-\infty$. Even more
intuitively, in a given background of $\varphi$, the second
scalar field $\psi$ behaves as if its kinetic term were
multiplied by $[1-\lambda(\partial_\mu \varphi)^2]$. If the
$\varphi$-background is chosen such that $\lambda(\partial_\mu
\varphi)^2$ be negative enough, then $\psi$ will behave as a
ghost, and its contribution to the Hamiltonian density will be
unbounded by below. Therefore, there do exist field
configurations such that $\mathcal{H}$ is as negative as one
wishes, and this proves the instability of the toy model. Such a
hand-waving argument can now also be used on more involved
models, for instance $\mathcal{L} = -(\partial_\mu \varphi)^2
-(\partial_\mu \psi)^2 + \lambda\Box\varphi(\partial_\mu
\psi)^2$, which looks a little more like Eq.~(\ref{e.nonminF2}),
where $\Box\varphi$ plays the r\^ole of the scalar curvature $R$,
involving second derivatives. The Hamiltonian analysis is now
much more involved, because the presence of third derivatives of
the fields in their equations implies the existence of new
excitations (and the standard Ostrogradski definition of
conjugate momenta cannot be followed because we are in a
degenerate case). But it is still clear that in a given
background where $\lambda\Box\varphi$ is large enough, then
$\psi$ behaves as a ghost and can make the Hamiltonian density
tend towards $-\infty$. Now, if we try to apply this argument to
action (\ref{e.nonminF2}) itself, we understand that we need to
consider large enough (positive or negative) curvatures $R$ such
that the nonminimal coupling $R F^2$ could change the global sign
of the vector kinetic term. Particular case might thus be safe,
for instance if one needs to be in the interior of a black hole
horizon to reach such a condition. Moreover, one may devise
models such that the function $\Xi(x) = \xi_0 + 0 x + \xi_2
x^2+\dots$ does not contain any linear term. Therefore, the above
hand-waving argument does not \textit{prove} that all models
(\ref{e.nonminF2}) are unstable, although we do expect so because
of the presence of higher derivatives in their field equations.
We will anyway disregard this class of models, because such
higher derivatives mean that they involve extra degrees of
freedom, in addition to the single spin-1 and spin-2 fields we
wished to consider.

\section{Dimensional reduction of Lovelock invariants and cosmological
phenomenology}\label{sec5}
A Lovelock invariant is defined in even dimension $D$ as a
Lagrangian density proportional to
$L_D \equiv \varepsilon^{\mu_1\mu_2\dots\mu_D}
\varepsilon^{\nu_1\nu_2\dots\nu_D}
R_{\mu_1\mu_2\nu_1\nu_2}
R_{\mu_3\mu_4\nu_3\nu_4}\dots
R_{\mu_{D-1}\mu_D\nu_{D-1}\nu_D}$, involving thus a product of
$D/2$ Riemann curvature tensors. The best known examples are the
cosmological constant $\Lambda$ corresponding to $D=0$, the
Einstein-Hilbert Lagrangian $R$ corresponding to $D = 2$, and the
Gauss-Bonnet density $R_{\mu\nu\rho\sigma}^2-4R_{\mu\nu}^2+R^2$
corresponding to $D = 4$. The integral of $L_D$ over a
$D$-dimensional spacetime gives a number depending only on the
topology, therefore its variational derivative vanishes and it
does not contribute to the field equations. In dimensions lower
than $D$, the density $L_D$ vanishes identically. On the other
hand, $L_D$ defines a nontrivial dynamics when considered in
dimensions higher than $D$ (like $R$ or $\Lambda$ in 4
dimensions). But in spite of the presence of several Riemann
tensors (for $L_{D\geq4}$), each of them involving second
derivatives of the metric, the corresponding field equations
remain of second order. Indeed, any third (or higher) derivative
must appear in a form similar to
$R_{\mu_1\mu_2\nu_1\nu_2;\mu_3}$, multiplied by the antisymmetric
Levi-Civita tensor $\varepsilon^{\mu_1\mu_2\dots\mu_D}$, and
therefore vanishes by virtue of the Bianchi identity
$R_{\nu_1\nu_2[\mu_1\mu_2;\mu_3]} = 0$. The absence of
higher-order derivatives in the field equations does not
guarantee the stability of the corresponding models, but it
proves at least that no extra degree of freedom is excited, and
that the generic ghost modes of higher-order theories are
avoided. If the Gauss-Bonnet density
$R_{\mu\nu\rho\sigma}^2-4R_{\mu\nu}^2+R^2$ is considered in 5
dimensions, for instance, it does contribute to the field
equations, but keeping them of second order. When performing a
Kaluza-Klein dimensional reduction, where $g_{\mu5}$ is
interpreted as a vector field $A_\mu$ in four dimensions, we thus
get a nontrivial vector-curvature coupling which does not
generate higher-order field equations, and avoids thus the deadly
instabilities caused by ghost modes\footnote{Moreover, the
dimensional reduction of Lovelock invariants always generates
(gauge-invariant) combinations of the Faraday tensor
$F_{\mu\nu}$, therefore the ghostlike mode $A_0$ is never excited
either; see Sec.~\ref{secIIA}.}. We will analyze below the
cosmology generated by such a coupling. Similar models can be
constructed by considering the dimensional reduction of
higher-order Lovelock invariants $L_6, L_8, \dots$, and even more
general vector models coupled to both curvature and scalar fields
are obtained by dimensionally reducing the so-called ``Galileon''
models recently introduced in Ref.~\cite{galileon1} and
generalized in curved spacetimes in Refs.~\cite{galileon2}. As we
will see below, even the simplest case of a dimensionally-reduced
Gauss-Bonnet density $L_4$ suffices to generate an interesting
cosmological evolution for the vector field.

\subsection{Nonminimal couplings to the Riemann tensor}
We consider the class of models
\begin{eqnarray}\label{action15}
S&=&\int\mathrm{d}^4x\sqrt{-g}\left[\frac{R}{2\kappa}
-\frac{1}{4}F^2 +\frac{1}{4}\xi R F^2
+\frac{1}{2}\eta R_{\mu\nu}F^{\mu\rho}{F^\nu}_\rho\right.
\nonumber\\
&&\left. +\frac{1}{4}\zeta R_{\mu\nu\rho\sigma}F^{\mu\nu}F^{\rho\sigma}
\right]
+ S_{\rm matter}[\psi_m;g_{\mu\nu}],
\end{eqnarray}
with the same notation as in the previous sections, and where
$\xi$, $\eta$ and $\zeta$ are constant parameters. Such theories
lead to generalization of the Maxwell theory that imply a
variable speed of light~\cite{cvar1,cvar2,r.cestc,teyssan} (i.e.
propagation velocity of the vector field if identified to the one
describing the photon~\cite{r.eu}). However, as shown in
\cite{r.horndeski}, the corresponding field equations are of
second order if and only if the parameters $\xi$, $\eta$ and
$\zeta$ satisfy
\begin{equation}\label{e.cond}
\eta+2\xi = 0,\qquad
\zeta=\xi,
\end{equation}
and this is precisely what is obtained by dimensionally reducing
the Gauss-Bonnet density $L_4$ written in a 5-dimensional
spacetime~\cite{r.gb1,r.gb2}. This can also be checked explicitly
by deriving the vector field equations
\begin{eqnarray}\label{e.MaxLov}
&&(1-\xi R)F^{\mu\nu}_{\hphantom{\mu\nu};\nu}
-\eta\left(R^\mu_\lambda F^{\lambda\nu}_{\hphantom{\lambda\nu};\nu}
- R^\nu_\lambda F^{\lambda\mu}_{\hphantom{\lambda\nu};\nu}
\right)\nonumber\\
&&-\zeta R^{\mu\nu\rho\sigma} F_{\rho\sigma;\nu}
-\frac{1}{2}(2\xi+\eta) R_{,\nu}F^{\mu\nu}\nonumber\\
&&-(\eta+2\zeta)R^\mu_{\lambda;\nu} F^{\lambda\nu} =0,
\end{eqnarray}
in which third derivatives of the metric occur (in the form of
first derivatives of the curvature tensor) unless relations
(\ref{e.cond}) are satisfied. Similarly, the Einstein equations
involve third derivatives of the vector $A_\mu$ unless
(\ref{e.cond}) are satisfied. Since such higher derivatives would
excite new, generically ghostlike, degrees of freedom, implying
the instability of the model, we will restrict our study to the
particular case (\ref{e.cond}). However, second-order field
equations do not suffice to warrant the consistency of the model.
These equations should also be hyperbolic, and the corresponding
Hamiltonian should be bounded by below. We will not perform here
this analysis, because it is even more complex than in the case
of couplings like $R_{\mu\nu}A^\mu A^\nu$. However, we wish to
emphasize that this particular class of models offers an
interesting phenomenology for cosmology and should thus deserve
more attention.

\subsection{Cosmological dynamics}

Consider action~(\ref{action15}) where the matter fields reduce
e.g. to a single scalar field $\phi$ evolving in a potential
$v(\phi)$ that is assumed to drive an inflationary phase in the
early universe. We consider the vector field as a test field
whose evolution is then given by Eq.~(\ref{e.MaxLov}). Using the
same notation (\ref{eq:AinTermsOfB}) as in Sec.~\ref{secCD1}
above, homogeneity in a Friedmann-Lema\^{\i}tre spacetime with
metric~(\ref{FLmetric}) implies $\partial_i A_\mu = 0$, so that
the only nonvanishing component of the Faraday tensor is
\begin{equation}
F_{0i} = a(\dot B_i + HB_i).
\end{equation}
We recall that the Weyl tensor of a Friedmann-Lema\^{\i}tre
spacetime strictly vanishes, $C_{\mu\nu\rho\sigma}=0$, and that
(restricting to a spatially Euclidean spacetime) the nonvanishing
components of the Ricci tensor are given by
\begin{equation}
R^0_0 = 3(\dot H+H^2),\qquad
R^i_j = (\dot H+ 3 H^2)\delta^i_j,
\end{equation}
so that the only nonvanishing components of the Riemann tensor
are (see e.g. Ref.~\cite{livrecosmo})
\begin{equation}
{R^i}_{jml}= a^2 H^2(\delta^i_m\gamma_{jl}-\delta^i_l\gamma_{jm}),\quad
{R^0}_{i0j} = a^2 (\dot H+H^2) \gamma_{ij}.
\end{equation}
The evolution equation for $B_i$ then reduces to
\begin{eqnarray}
&&\Bigl[1 - 6\xi (\dot H + 2 H^2)
-\eta (4 \dot H + 6 H^2)-2\zeta(\dot H+H^2)\Bigr]
\nonumber\\
&&\times (\dot F^{i0}+3HF^{i0})
+\Bigl[-(6\xi +4\eta +2\zeta)(\ddot H+4\dot H H)
\nonumber\\
&&+4(\eta+\zeta)\dot H H \Bigr]F^{i0}=0.
\end{eqnarray}
Restricting to the conditions~(\ref{e.cond}), it leads to the equation
\begin{eqnarray}\label{e.MaxLOVFL}
&&(1+\eta H^2)\ddot B_i +
3\left[1+\eta \left(\frac{2}{3}\dot H+H^2\right)\right]
H \dot B_i\nonumber\\
&&+\left[\left(1+3\eta H^2\right)\dot H
+ 2\left(1+\eta H^2\right)H^2\right]B_i =0.\quad
\label{eq:Bi}
\end{eqnarray}

Let us first assume that the universe is undergoing a slow-roll
inflationary phase close to a de Sitter phase, so that we can
assume $H\sim$ const. and $-\dot H/H^2 = \varepsilon \ll~1$
($\varepsilon>0$ in most slow-roll inflationary models). If the
parameter $\eta$ is chosen to be negative, then a fine-tuned
value $H^2 \approx -(1+\varepsilon)/\eta$ is such that
Eq.~(\ref{eq:Bi}) reads $\ddot B_i + (1-2\varepsilon) H \dot B_i
-3\varepsilon H^2 B_i = 0$, and therefore does not involve any
undifferentiated $B_i$ at lowest order in $\varepsilon$ [this can
easily be made exact thanks to an even finer tuning of $H(t)$].
The two solutions of this equation are thus a decaying mode $B_i
\sim \exp[-Ht] \sim 1/a$ and an almost constant one $B_i \sim
\exp[3\varepsilon Ht] \sim a^{3\varepsilon}$ ---~even slightly
increasing if $\varepsilon > 0$. It follows that a slow-rolling
vector field can survive the expansion, contrary to the standard
lore on vector fields, but at the price of a fine tuning of the
expansion rate $H$, related to the nonminimal vector-gravity
coupling constant $\eta$.

Let us also consider the dynamics of the vector field assuming
the background dynamics is given by $a(t)\propto t^p$ ($p=1/2$
for a radiation-dominated universe, $p=2/3$ for a
matter-dominated universe, and the limit $p\gg1$ corresponds to a
power-law inflationary model with $\varepsilon=1/p$).
Equation~(\ref{eq:Bi}) then reduces to
\begin{eqnarray}\label{eq-pdyn}
&&(1+\eta H^2)\ddot B_i
+ \left[3 +\left(3-\frac{2}{p}\right)\eta H^2\right]
H\dot B_i\nonumber\\
&&+\left[\left(2-\frac{1}{p}\right)
+\left(2-\frac{3}{p}\right)\eta H^2\right] H^2 B_i = 0.\quad
\end{eqnarray}
In the case of inflation, we find again that the field is diluted
unless one imposes the previous fine tuning $1+\eta H^2=-1/p$,
which requires $\eta <0$.

To discuss the dynamics during the matter and radiation-dominated
era, let us introduce the time scale $\tau_*=p\sqrt{|\eta|}$. In
the radiation era, the coefficient of $B_i$ is always
proportional to $\eta$ (instead of being zero in the standard
case). At early times ($t\ll\tau_*$), Eq.~(\ref{eq-pdyn}) reduces
to $\ddot B_i -\dot B_i/(2t) -B_i/t^2=0$ which has two solutions,
a decaying mode $\propto1/\sqrt{t}\propto a^{-1}$ and a growing
mode $\propto t^2\propto a^4$ while, $C_i\equiv \dot B_i + H B_i$
behaves as $C_i\propto t\propto a^2$. At later times
($t\gg\tau_*$), Eq.~(\ref{eq-pdyn}) reduces to $\ddot B_i +3\dot
B_i/(2t) -\eta B_i/(4t^4)=0$, which differs from the standard
equation by the term proportional to $B_i$. The solutions of such
an equation are given in terms of Bessel functions and will be
oscillating if $\eta<0$ while they have a mode $\propto
(t/\tau_*)^{-1/4}K_{1/4}(\tau_*/t)$ if $\eta>0$ that grows and
then freezes to a constant. These behaviors at early and late
times differ from the standard dynamics of a vector field and
exist whatever the value of $\eta$. In the matter era, the
dynamics is only modified at early times ($t\ll\tau_*$) since
Eq.~(\ref{eq-pdyn}) reduces to $\ddot B_i +9 t \dot B_i/(2\eta)
-10B_i/(9 t^2)=0$, the main modification arising from the fact
that the coefficient of $\dot B_i$ is now proportional to $1/H$
and not to $H$ anymore. This equation has a growing mode.

\subsection{Discussion}

In this class of theories, a slow-rolling vector field can
survive during inflation, contrary to the standard lore on vector
fields, but at the price of a fine tuning of the expansion rate
$H$, related to the nonminimal vector-gravity coupling constant
$\eta$. It requires that $\eta$ be negative and is related to the
energy scale of inflation by $|\eta|\sim1/H_{inf}^2$. The general
action should thus contain terms of the form
\begin{equation}
\mathcal{L}\supset \frac{1}{2}M_p^2R
- \frac{1}{4}F^2 -\frac{1}{2} \eta H^2 F_{0i}^2,
\nonumber
\end{equation}
where $M_p$ is the Planck mass. Then, since during inflation
$R\sim12 H_{inf}^2$ while we need $\eta H_{inf}^2\sim-1$, the
correction term to the standard Einstein-Maxwell Lagrangian is of
the order of $F_{0i}^2/2=-F^2/4$. As long as $H_{inf}/M_p<1$, as
is usually the case in inflation, this correction is negligible
compared to the Einstein-Hilbert while being of the same order as
the Maxwell term. In spirit, this solution leading to a
slow-rolling vector field is similar to the one invoked in
Refs.~\cite{golovnev,kanno}, which used a coupling of the form
$\xi R A^2$, that we saw to be unstable. We cannot prove at this
stage that this will not be the fate of this model that needs to
be analyzed in detail.

We have also seen that the dynamics during the radiation and
matter-dominated eras allows for growing solutions whatever the
value of the parameter $\eta$. This opens an interesting
phenomenology that we postpone to further study.

\section{Conclusions}
In this article, we have investigated general models of vector
fields that have recently been considered in cosmology, in
relation with a source of anisotropy or the construction of
MOND-inspired field theories.

We have shown that the class of $f(F^2)$-theories suffers from
hyperbolicity problems, while both $f(F^2)$ and $f(F^2,F\tilde
F)$ models predict a dilution of the vector field during the
cosmological expansion.

When allowing for a nonminimal coupling to the metric, we have
proven that the class of $f(A^2)R$-theories has a Hamiltonian
which is unbounded from below, while the $f(F^2)R$-models involve
higher derivatives of the fields and thus contain extra degrees
of freedom (which are generically expected to carry negative
energy).

These results set strong constraints on vector field models, as
long as they are considered as fundamental theories ---~i.e.,
notably, that no field entering the action is considered as a
fixed background that cannot be varied. [From a theoretical point
of view, let us remind that an action is not just a list of
symbols but involves also the definitions of these symbols, e.g.
what are the fundamental fields; see the discussion of the
difference between $A_\mu$ and $\partial_\mu\phi$, or the
difference between a potential and a Lagrange multiplier.] But
even as effective models, the constraints we derived for their
stability and causality should always be satisfied in their
domain of validity, and at least in the domain where their
cosmological evolution is studied. It happens that to avoid the
dilution of the vector field during the expansion of the
Universe, one would need the nonlinear terms to be of the same
order of magnitude as the main kinetic term, i.e., precisely in
conditions where the positivity of the Hamiltonian and the
well-posedness of the Cauchy problem should be checked carefully.

To finish, we pointed out that in the class of theories obtained
by dimensional reduction of Lovelock invariants, there exist
cases that allow for the existence of a slow-rolling vector
field. Although we did not study the boundedness by below of the
Hamiltonian nor the hyperbolicity of the field equations, because
of their complexity, we underlined that the field equations
remain of second order in spite of the nonminimal coupling of the
vector field to curvature. Such models contain thus only the
spin-1 and spin-2 degrees of freedom we wished to consider (in
addition to other matter fields), and they are phenomenologically
quite appealing for cosmology.

Similar models as the one we studied in Sec.~\ref{sec5} are
obtained by dimensionally reducing higher-order Lovelock
invariants, and more general tensor-vector-scalar models yielding
second-order field equations can also be defined by dimensional
reduction of Galileon actions~\cite{galileon1,galileon2} written
in more than 4 dimensions. It is also possible that the
construction of scalar Galileons can be generalized to vector
fields, yielding nonminimal vector-curvature couplings of a
different nature than those obtained from Lovelock invariants.
All such second-order models deserve being studied both
mathematically and for their phenomenological predictions
in a cosmological context.

\begin{acknowledgments}
C.P. would like to thank Institut d'Astrophysique de Paris
for its kind hospitality during part of this project.
\end{acknowledgments}


\end{document}